\documentclass[pra,english,reprint, longbibliography, superscriptaddress, breaklinks=true, showkeys, showpacs=false, nofootinbib]{revtex4-2}

\usepackage[T1]{fontenc}
\usepackage[utf8]{inputenc}
\usepackage{color}
\usepackage{babel}
\usepackage{amsmath}
\usepackage{amssymb}
\usepackage{subfigure}
\usepackage{graphicx}
\usepackage{physics} 
\usepackage{dsfont}
\usepackage{hyperref}
\usepackage{float}
\newcommand{\integral}{\displaystyle\int}
\newcommand{\vX}{\textbf{X}}
\newcommand{\vx}{\textbf{x}}

\begin{document}

\title{Work distribution of quantum fields in static curved spacetimes}

\author{Rafael L. S. Costa\href{https://orcid.org/0009-0008-5735-5917}{\includegraphics[scale=0.05]{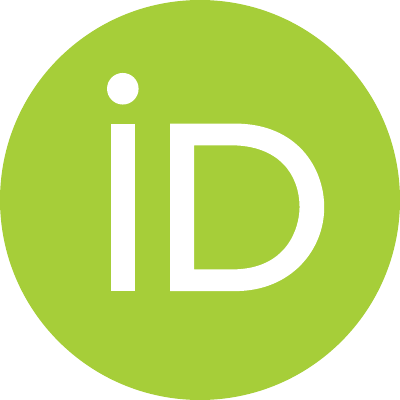}}}
\email{dolago.rafael@gmail.com}
\affiliation{QPequi Group, Institute of Physics, Federal University of Goi\'as, Goi\^ania, Goi\'as, 74.690-900, Brazil}

\author{Marcos L. W. Basso\href{https://orcid.org/0000-0001-5456-7772}{\includegraphics[scale=0.05]{orcidid.pdf}}}
\email{marcoslwbasso@hotmail.com}
\affiliation{Department of Applied Mathematics, Universidade Estadual de Campinas, 13083-859 Campinas, S\~ao Paulo, Brazil}

\author{Jonas Maziero\href{https://orcid.org/0000-0002-2872-986X}{\includegraphics[scale=0.05]{orcidid.pdf}}}
\email{jonas.maziero@ufsm.br}
\affiliation{Physics Department, 
Federal University of Santa Maria, 97105-900,
Santa Maria, RS, Brazil}

\author{Lucas C. C\'eleri\href{https://orcid.org/0000-0001-5120-8176}{\includegraphics[scale=0.05]{orcidid.pdf}}}
\email{lucas@qpequi.com}
\affiliation{QPequi Group, Institute of Physics, Federal University of Goi\'as, Goi\^ania, Goi\'as, 74.690-900, Brazil}

\begin{abstract}
We investigate the formulation of work distributions for quantum scalar fields in static curved spacetimes by extending the Ramsey interferometric protocol originally developed in previous works for flat spacetimes. The use of Unruh–DeWitt particle detectors provides a causally consistent framework to define and measure work statistics, avoiding the limitations of the two-time projective measurement scheme in relativistic quantum field theory. We derive a non-perturbative expression for the characteristic function of the quantum field and apply it to thermal Kubo–Martin–Schwinger (KMS) states, showing that the resulting work distributions satisfy both the Crooks fluctuation theorem and the Jarzynski equality. Furthermore, we analyze the case of a pointlike detector, obtaining compact expressions for the first two moments of the work distribution, allowing us to recover the standard fluctuation-dissipation relation in the high-temperature limit. Our results demonstrate that fluctuation theorems hold for quantum fields interacting with Unruh–DeWitt particle detectors in static curved spacetimes.
\end{abstract}


\maketitle
 
%%%%%%%%%%%%%%%%%%%%%%%%%%%%%%%%%%%%%%%%%%%%%%%%%%%%%%%%%%%%
%%%%%%%%%%%%%%%%%%%%%%%%%%%%%%%%%%%%%%%%%%%%%%%%%%%%%%%%%%%%
\section{Introduction} 

The intersection of quantum field theory and thermodynamics in curved spacetime has revealed deep and surprising connections between energy flow, horizon dynamics, and information. A landmark result in this direction is Hawking’s prediction that black holes emit thermal radiation due to quantum effects near the event horizon~\cite{Hawking1975}, complementing the thermodynamic analogy previously established by Bekenstein~\cite{Bekenstein1973}. These insights support the now well-established framework of quantum field theory in curved spacetimes, where particle creation, vacuum structure, and temperature become observer-dependent phenomena~\cite{Birrell1982,Wald1994}.

In parallel, the field of nonequilibrium statistical mechanics has developed powerful frameworks to understand irreversible processes from first principles. Central among these are fluctuation theorems, which describe the precise statistical properties of entropy production, work, and heat in systems driven arbitrarily far from equilibrium. The Crooks fluctuation theorem~\cite{Crooks1999} and the Jarzynski equality~\cite{Jarzynski1997} stand among the most prominent results. In the quantum regime, fluctuation theorems have been extended using time-ordered correlation functions~\cite{Talkner2007} or interferometric techniques~\cite{Dorner2013,Mazzola2013}, and have also been generalized to open quantum systems~\cite{Campisi2011,Esposito2009} and for generalized measurements and dynamics~\cite{Kafri2012,Rastegin2014}.

Bringing these two directions together, the formulation of fluctuation theorems for quantum fields in curved spacetimes remains largely unexplored. This is a nontrivial task: the lack of global symmetries and the absence of a preferred vacuum or time parameter in general curved backgrounds forbids the direct application of standard fluctuation relations. Furthermore, the conventional two-time measurement protocol used in quantum thermodynamics is generally incompatible with relativistic causality~\cite{Sorkin1993,Borsten,Bostelmann,Anastopoulos2022}, necessitating more sophisticated measurement frameworks. However, recent advances in relativistic quantum information theory and operational quantum field theory suggest that localized, causally consistent measurement models --- such as those based on Unruh-DeWitt (UDW) particle detectors~\cite{Unruh1976,DeWitt1979,Takagi1986,Schlicht04,Louko06,Louko08,Satz2007,Ramon21,Tjoa22,Ramon23,Polo,Perche2024} --- can restore operational clarity while preserving relativistic consistency.

Several works have already investigated the thermal properties of quantum field theories in flat and curved spacetimes~\cite{Haag1996,Sewell1982,Takagi1986,Kay1991}, as well as the thermalization of UDW particle detectors~\cite{Fewster,Garay,Aubry,Perche2021}. Moreover, a well-defined notion of work distribution involving quantum fields in Minkowski spacetime was established by Ortega~\textit{et al.}~\cite{Ortega} through the use of Ramsey interferometry~\cite{Dorner2013,Mazzola2013}. A further work~\cite{Bonfill} showed that the notion of work distribution established in Ref.~\cite{Ortega} satisfies the first law of thermodynamics up to second moments.

Regarding non-equilibrium thermodynamics in curved spacetimes, some notable developments have already been achieved. In the context of linear phenomena, Mottola~\cite{Mottola} derived a fluctuation-dissipation relation adapted to curved backgrounds. Iso \textit{et al.}~\cite{Iso2011} investigated non-equilibrium fluctuations at black hole horizons by employing Jarzynski equality together with the generalized second law of thermodynamics~\cite{Bekenstein74,Wall2012} and a fluctuation relation was obtained for a quantum field theory model in an expanding Universe scenario in Ref.~\cite{Liu2016}. Moreover, a fully general relativistic quantum fluctuation theorem based on the two-point measurement scheme for a localized nonrelativistic quantum system was presented in Ref.~\cite{Basso2025}, extending the results obtained in Ref.~\cite{Basso2023}. Similarly, a general relativistic fluctuation theorem was obtained in Ref.~\cite{Cai25} for stochastic classical systems. Despite all these achievements, a detailed fluctuation theorem applicable to quantum fields interacting with localized apparatuses, or particle detectors, in curved spacetimes has not yet been fully developed.

Building upon the aforementioned contributions, we take a step further by extending the protocol defined in Ref.~\cite{Ortega} to construct the characteristic function and the Ramsey-scheme work distribution for a quantum scalar field in a globally hyperbolic and static curved spacetime. Then, we derive a non-perturbative expression for the characteristic function of the quantum field. By non-perturbative, we mean that the unitary time evolution in the interaction picture, generated by the detector–field interaction Hamiltonian, can be expressed as a finite sum of bounded operators. In this sense, no truncation of the Dyson series at a finite order in the coupling strength is performed, in contrast to the usual practice in weak-coupling scenarios. This, in turn, allows us to obtain a closed expression for the characteristic function, which we then apply to Kubo–Martin–Schwinger (KMS) thermal states, showing that the ratio of forward and reverse work probability distributions satisfies both the detailed Crooks theorem and the Jarzynski equality, thereby extending fluctuation theorems to relativistic quantum fields. Furthermore, by analyzing a pointlike detector, we derive simple expressions for the first two moments of the work distribution, recovering the standard fluctuation-dissipation relation in the high-temperature regime.

The article is structured as follows. In Sec.~\ref{sec:II}, we start by reviewing the quantization of scalar fields in static spacetimes, together with the construction of Fermi normal coordinates and the formulation of UDW particle detectors. After that, we extend the approach proposed in Ref.~\cite{Ortega} to introduce the notion of work distribution for a quantum field in static curved spacetimes. In Sec.~\ref{sec:KMSwork}, we present the non-perturbative expression for the characteristic function of the quantum field and show that the Ramsey scheme probability distribution satisfies the fluctuation theorems. In addition, the case of a pointlike detector is also discussed. Finally, our conclusions are presented in Sec.~\ref{conclusions}. We employ the metric signature ($-,+,+,+$) and natural units throughout the article. Also, Greek indices run from 0 to 3 while Latin ones stand for the spatial coordinates and run from 1 to 3.

%%%%%%%%%%%%%%%%%%%%%%%%%%%%%%%%%%%%%%%%%%%%%%%%%%%%%%%%%%%%
%%%%%%%%%%%%%%%%%%%%%%%%%%%%%%%%%%%%%%%%%%%%%%%%%%%%%%%%%%%%
\section{Work field distribution in static curved spacetimes}
\label{sec:II}

In this section, we briefly review the quantization of scalar fields in static spacetimes, Fermi normal coordinates, and UDW particle detectors. The goal is to make the article more self-contained and to establish the notation that will be used in what follows. We then extend the procedure introduced in Ref.~\cite{Ortega} to define the work distribution for a quantum field in static curved spacetimes.

%%%%%%%%%%%%%%%%%%%%%%%%%%%%%%%%%%%%%%%%%%%%%%%%%%%%%%%%%%%%
\subsection{Field quantization in static spacetimes}
\label{sec:field_q}

Let us begin by considering the Klein-Gordon action for a massive real scalar field $\phi$ propagating through a curved spacetime background:
\begin{equation}
S[\phi] \equiv -\dfrac{1}{2}\int_{\mathcal{M}} \dd^4 \mathsf{x} \sqrt{-g}\left[g^{\mu\nu}\nabla_{\mu}\phi\nabla_{\nu}\phi + m^2\phi^2\right],
\label{kgaction}
\end{equation}
where the integration extends over the entire spacetime manifold $\mathcal{M}$ equipped with a metric tensor $g_{\mu\nu}$, with $\sqrt{-g}\dd^4\mathsf{x}$ denoting the invariant volume element and $\nabla_\mu$ representing the covariant derivative compatible with the metric. Moreover, $\mathsf{x}$ denotes an arbitrary point in $\mathcal{M}$, which may be parameterized by a suitable coordinate system.

Variation of the above action with respect to the metric tensor leads us to the energy-momentum tensor, which encodes both the field's stress-energy content and its coupling to gravity, expressed as
\begin{equation}
T_{\mu\nu} = \nabla_{\mu}\phi\nabla_{\nu}\phi - \frac{1}{2}g_{\mu\nu}\left[\nabla_{\alpha}\phi\nabla^{\alpha}\phi + m^2\phi^2\right],
\label{energymomentum}
\end{equation}
while the variation with respect to the scalar field $\phi$ leads us to the Klein-Gordon equation
\begin{align}
    (\nabla_\mu \nabla^{\mu} - m^2) \phi = 0. \label{eq:KG}
\end{align}

The quantization procedure for a real scalar field $\phi$ requires the field to be treated as an operator-valued distribution, $f \mapsto \hat{\phi}(f)$, for $f \in C^{\infty}_0(\mathcal{M})$ (or a suitable test-function space), acting densely on a Hilbert space $\mathcal{H}$ of one-particle states, such that the map $f \mapsto \hat{\phi}(f)$ is linear with 
\begin{equation}
 \hat{\phi}(f) \equiv \int_{\mathcal{M}} \dd^4 \mathsf{x} \sqrt{-g} f(\mathsf{x}) \hat{\phi}(\mathsf{x}).    
\end{equation}
The field is self-adjoint, i.e., $\hat{\phi}(f)^\dagger = \hat{\phi}(f)$; the Klein-Gordon equation (given by Eq.~\eqref{eq:KG}) is satisfied. Finally, the canonical commutation relations are fulfilled, i.e., for all $f_1, f_2 \in C^{\infty}_0(\mathcal{M})$, we have 
\begin{equation}
[\hat{\phi}(f_1), \hat{\phi}(f_2)] = -i \Delta(f_1, f_2) \hat{\mathbb{I}},
\end{equation}
where $\hat{\mathbb{I}}$ denotes the identity operator and $\Delta(f_1,f_2)$ is the smeared causal propagator defined as
\begin{align}
    \Delta(f_1, f_2) \equiv \int_{\mathcal{M}} \dd^4 \mathsf{x} \sqrt{-g} f_1(\mathsf{x}) E f_2(\mathsf{x}), \label{eq:Delta}
\end{align}
with $E$ being the advanced-minus-retarded Green operator associated with the Klein–Gordon operator $\nabla_\mu \nabla^\mu - m^2$ through
\begin{align}
    Ef(\mathsf{x}) = \int_{\mathcal{M}} \dd^4 \mathsf{x'}\sqrt{-g'}\left(G^{adv}(\mathsf{x},\mathsf{x}') - G^{ret}(\mathsf{x},\mathsf{x}')\right)f(\mathsf{x}'). 
\end{align}

Since we consider a globally hyperbolic static spacetime, where the spacetime admits a timelike Killing vector field $\mathcal{X}^\mu = (\partial_t)^\mu$ orthogonal to the hypersurfaces $\Sigma_t: t = \text{constant}$, the line element can be written as $\dd s^2 = -N^2(\vx) \dd t^2 + h_{ij}(\vx) \dd x^i \dd x^j$, with $h_{ij}$ being the Riemannian metric on each hypersurface $\Sigma_t$, and $\vx$ denotes the spatial coordinates on $\Sigma_t$. In this case, the Klein–Gordon equation takes the form $(-\partial_t^2 - K) \hat{\phi} = 0$, where $K = N^2 \left((1/\sqrt{-g})\partial_i (\sqrt{-g} \partial^i) + m^2 \right)$ is a positive self-adjoint spatial operator. This leads to the spectral decomposition $K F_j = \omega_j^2 F_j$, where $F_j(\vx)$ are the spatial mode functions satisfying the orthonormality condition $\int_{\Sigma_t} \dd^3 \vx\sqrt{-g}N^{-2}F_j F_{j'}^* = \delta_{\xi}(j-j')$. This spectral decomposition allows us to expand the unsmeared field operator as 
\begin{equation}
    \hat{\phi}(\mathsf{x}(t,\vx)) = \int \dfrac{\dd\xi(j)}{\sqrt{2\omega_j}}\left[F_j(\vx)e^{-i\omega_j t}\hat{a}_{j} + F^*_j(\vx)e^{i\omega_j t}\hat{a}^{\dagger}_{j}\right], \label{eq:phidec}
\end{equation}
where $\hat{a}^{\dagger}_j$ and $\hat{a}_j$ are the creation and annihilation operators, respectively, with $[\hat{a}_{j},\hat{a}^{\dagger}_{j'}] = \delta_{\xi}(j - j')\hat{\mathbb{I}}$. The zero-temperature vacuum state of the theory is defined by $\hat{a}_{j} | \Omega_{\infty}\rangle = 0$, and the associated bosonic Fock space can be constructed from the one-particle Hilbert space $\mathcal{H}$ in the standard way.

The Hamiltonian of the free quantum field is defined with respect to the timelike Killing vector of spacetime as $\hat{H}_{\phi} \equiv \int_{\Sigma_t} \dd^3 \vx \sqrt{h} n^\mu \mathcal{X}^\nu T_{\mu\nu}(\hat{\phi},\hat{\phi})$, with $n^{\mu} = \mathcal{X}^\mu/\sqrt{- \mathcal{X}_\nu \mathcal{X}^\nu}$.

To describe thermal equilibrium at nonzero temperature $T = 1/\beta$, one introduces KMS states~\cite{Kubo1957,Martin1959,Sanders10}, denoted by $\hat{\rho}_{\beta}$, which replace the pure vacuum by quasifree states satisfying the KMS condition with respect to the modular flow generated by the Hamiltonian $\hat{H}_{\phi}$~(or, equivalently, with respect to the time-translation symmetry generated by $\mathcal{X}^\mu$). Hence, the KMS condition provides a rigorous and general notion of thermality in this context, which can be formulated in terms of the two-point Wightman function, $\mathcal{W}_{\beta}(t,t') \equiv \mathrm{Tr}\{\hat{\phi}(t,\vx)\hat{\phi}(t',\vx')\hat{\rho}_\beta\} \equiv \langle \hat{\phi}(t,\vx)\hat{\phi}(t',\vx')\rangle_{\beta}$, through the relation $\mathcal{W}_{\beta}(t - i\beta, t') = \mathcal{W}_{\beta}(t',t)$. Since the Wightman function is invariant under time translation, i.e., $\mathcal{W}_{\beta}(t,t') = \mathcal{W}_{\beta}(t - t' \equiv \Delta t)$, the KMS condition becomes $\mathcal{W}_{\beta}(\Delta t - i\beta) = \mathcal{W}_{\beta}(- \Delta t)$.

%%%%%%%%%%%%%%%%%%%%%%%%%%%%%%%%%%%%%%%%%%%%%%%%%%%%%%%%%%%%
%%%%%%%%%%%%%%%%%%%%%%%%%%%%%%%%%%%%%%%%%%%%%%%%%%%%%%%%%%%%
\subsection{Fermi normal coordinates and UDW particle detectors}
\label{sec:fermi}

In quantum field theory in curved spacetimes, the formulation of observables and detectors is inherently local and often tied to the experience of a family of observers. Here, we will consider that the detector is spatially smeared around a worldline segment, which represents the worldline of our static observer $\mathcal{O}$. So, let $\mathsf{x}(\tau) = \gamma(\tau)$ be the timelike trajectory of $\mathcal{O}$, which follows the orbits of the Killing vector field $\mathcal{X}^{\mu}$, in the spacetime $\mathcal{M}$ parametrized by proper time $\tau$, with four-velocity $u^\mu$ such that $u_{\mu} u^{\mu} = -1$. Hence
\begin{align}
    u^{\mu} =  \frac{\mathcal{X}^\mu}{\sqrt{- \mathcal{X}_\nu \mathcal{X}^\nu}} = \frac{\delta^{\mu}_t}{N|_{\gamma(\tau)}},
\end{align}
and the time coordinates are related by the lapse function $N$, i.e., $\dd \tau/\dd t = N|_{\gamma(\tau)}.$
Let us notice that $N$ is constant along the worldline $\gamma(\tau)$. Moreover, given that the orbits of $\mathcal{X}^{\mu}$ are given by $\varphi_t(t_0, \vx_0) = (t_0 + t, \vx_0)$, then $\gamma(\tau) = (N^{-1}(\vx_0)(\tau_0 + \tau), \vx_0) = \varphi_{\tau}(\tau_0, \vx_0)$.

However, when the detector is spatially smeared along a finite segment of its worldline, it becomes essential to describe its behavior using geometrically adapted coordinates. Fermi normal coordinates (FNC) provide exactly such a framework: a coordinate system defined in a tubular neighborhood around a timelike worldline $\gamma(\tau)$, which incorporates both the spacetime curvature and the proper acceleration of the observer~\cite{Perche2021,perche2022localized}. Hence, for the worldline $\gamma(\tau)$, a local orthonormal frame $\{e^{\mu}_{(a)}\}_{a = 0}^3$, satisfying $g_{\mu \nu} e_{a}^{\mu} e_{b}^{\nu} = \eta_{ab}$, can be defined by means of Fermi-Walker transport 
\begin{equation}
\frac{D_{\text{FW}} e^\mu_{(a)}}{d\tau} = (a_\nu e^\nu_{(a)}) u^\mu - (u_\nu e^\nu_{(a)}) a^\mu,
\end{equation}
where $a^\mu = u^\nu \nabla_\nu u^\mu$ denotes the observer's proper acceleration and $e^{\mu}_{(0)}$ is identified with the four-velocity $u^{\mu}$. The FNC $(\tau, X^i)$ are then defined such that the spatial coordinates $X^i$ are assigned via spacelike geodesics orthogonal to $\gamma$, with proper length measured along the initial direction determined by $e^\mu_{(i)}$~\cite{manasse1963fermi,poisson2004relativist}. Moreover, it is worth mentioning that, inside the tubular neighborhood around the timelike curve $\gamma(\tau)$, the Fermi normal coordinates $(\tau, X^i)$ and the symmetry-adapted coordinates $(t, x^i)$ are related by a well-defined coordinate transformation. 

The metric can be expanded in terms of the FNC, taking the following form
\begin{align}
		g_{\tau\tau} &= -(1 + a_i(\tau) X^i)^2 - R_{\tau i\tau j} (\tau)X^iX^j + \mathcal{O}(X^3),  \nonumber\\
		g_{\tau i} &= -\dfrac{2}{3}R_{\tau jik}(\tau)X^jX^k + \mathcal{O}(X^3), \label{eq:gFNC}\\
		g_{ij} &= \delta_{ij} -\dfrac{1}{3}R_{ikjl}(\tau)X^kX^l + \mathcal{O}(X^3),  \nonumber
	\label{fnc}
\end{align}
where $a_i(\tau)$ and $R_{\mu \nu \alpha \beta} (\tau)$ represent, respectively, the 4-acceleration and components of the Riemann curvature tensor in the Fermi normal coordinates evaluated along $\gamma(\tau)$. The presence of the acceleration term in $g_{\tau\tau}$ accounts for inertial effects due to non-geodesic motion, while the curvature terms encode the local tidal gravitational effects. It is worth noting that, since our observer $\mathcal{O}$ follows one of the orbits of the Killing vector $\mathcal{X}^{\mu}$, both the 4-acceleration and the components of the Riemann curvature tensor are independent of the proper time $\tau$ and the coordinate time $t$, with $R_{\tau jik}(\tau) = 0$~\cite{Medved04}.

The UDW particle detector, modeled as a two-level system smeared along the worldline $\gamma$ of our observer $\mathcal{O}$, whose free Hamiltonian with respect to the proper time $\tau$ is given by $H_D = \Omega \hat{\sigma}_3/2$, where $\Omega > 0$ is the energy gap and $\hat{\sigma}_3 = |0 \rangle \langle 0| - |1\rangle \langle 1|$ is one of the Pauli matrices. 

The localized interaction of the particle detector with the scalar field $\hat{\phi}(\mathsf{x})$, in the interaction picture as seen by $\mathcal{O}$, is given by
\begin{equation}
	\hat{H}_I(\tau) = \lambda \chi(\tau)\integral_{\Sigma_\tau} \dd^3 \vX \sqrt{-g} \psi(\vX)\hat{\phi}(\mathsf{x}(\tau,\vX)), \label{eq:Hint}
\end{equation}
where $\psi(\vX)$ encodes the spatial smearing of the detector around the curve $\gamma(\tau)$ and the switching function $\chi(\tau)$ governs the temporal profile of the interaction and has strong support within a finite time interval $[\tau_0, \tau_1]$. Moreover, $\hat{\phi}(\mathsf{x}(\tau,\vX))$ denotes the pullback of the field around the worldline $\gamma (\tau)$ in Fermi normal coordinates.

Some observations about the interaction Hamiltonian~\eqref{eq:Hint} are in order. First, we assume the rigidity condition on the detector~\cite{Martinez20,Martinez21}. Specifically, this condition states that the spacetime smearing can be expressed as the product of a switching function and a spatial smearing function that is time-independent in the reference frame associated with the observer’s trajectory. Furthermore, it is worth noting that the interaction Hamiltonian in Eq.~\eqref{eq:Hint} is a particular case of those considered in Refs.~\cite{Aubry,Perche2021,perche2022localized}. However, it is precisely this case that reduces to the interaction Hamiltonian used in the Ramsey protocol of Ref.~\cite{Ortega} in the flat-spacetime limit, while still providing a fully covariant description of a smeared particle detector interacting with a quantum field~\cite{Martinez20,Martinez21}.

Finally, the unitary evolution operator governing the interaction between the quantum field and the particle detector, as described by $\mathcal{O}$, is given by
\begin{align}
    \hat{U}_I = \hat{\mathcal{T}} e^{-i \int_{-\infty}^{\infty} \dd \tau \hat{H}_I(\tau) } \label{eq:U_I}
\end{align}
where $\hat{\mathcal{T}}$ is the time-ordering operator.

%%%%%%%%%%%%%%%%%%%%%%%%%%%%%%%%%%%%%%%%%%%%%%%%%%%%%
%\begin{figure}[t]
%    \centering
 %   \includegraphics[scale=0.8]{geogebra-export.png}
  %  \caption{Fermi tube and Fermi bound around a timelike worldline $\gamma(\tau)$. \textcolor{red}{LCC: Esta figura é realmente necessária? E se for, é possível mudar a cor verde por azul, de preferência um azul pastel, que fica melhor.}}
   % \label{fig:ftube}
%\end{figure}
%%%%%%%%%%%%%%%%%%%%%%%%%%%%%%%%%%%%%%%%%%%%%%%%%%%%

%%%%%%%%%%%%%%%%%%%%%%%%%%%%%%%%%%%%%%%%%%%%%%%%%%%%%%%%%%%%
%%%%%%%%%%%%%%%%%%%%%%%%%%%%%%%%%%%%%%%%%%%%%%%%%%%%%%%%%%%%
\subsection{Ramsey interferometry and the work field distribution}
\label{sec:ramsey}

To quantify work distributions in quantum systems with finite degrees of freedom, the two-point measurement (TPM) scheme is frequently employed~\cite{Talkner2007,Campisi2011}. The protocol begins by preparing the system in a thermal state, $\hat{\rho}_{0} = e^{-\beta \hat{H}(0)}/Z_0$, where $Z_0$ is the initial canonical partition function. Following preparation, an initial projective energy measurement projects the state onto an eigenstate of the initial Hamiltonian $\hat{H}(0)$. The projection onto this subspace is performed by $\hat{\Pi}^{0}_n$, and occurs with probability $p_n = \text{Tr}[\hat{\rho}_{0} \hat{\Pi}^{0}_n]$. The system then evolves unitarily $
\hat{\rho}_n(\tau) = \hat{U}_{\tau,0}\hat{\Pi}^{0}_n \hat{U}^{\dagger}_{\tau,0}$. At time $\tau$, a final projective measurement (implemented by $\hat{\Pi}^{\tau}_m$) yields outcome $E^{\tau}_m$ with conditional probability $p_{m|n} = \text{Tr}[\hat{\Pi}^{\tau}_m \hat{\rho}_n(\tau)]$. From this protocol, the work as a stochastic variable can be defined as $W_{m,n} = E^{\tau}_m - E^{0}_n$ and the work probability distribution density of the forward process can be constructed as $P_{\text{fwd}}(W) = \sum_{m,n}p_{m,n}\delta\left[W - W_{k,l}\right]$, where $p_{m,n} = p_{n}p_{m|n}$ is the joint probability.

However, TPM scheme fails in relativistic QFT due to its reliance on instantaneous projective measurements. As noted in Ref.~\cite{Ortega}, Ramsey interferometry resolves this causality violation by eliminating intermediate projections. The protocol established in Ref.~\cite{Ortega} operates entirely within the unitary framework, replacing destructive measurements with coherent superpositions. It thereby preserves spacetime's causal structure while extracting equivalent thermodynamic information about the field.

The interferometric protocol is the following:
\begin{enumerate}
    \item The UDW particle detector is prepared in its ground state $\ket{0}$, after which a Hadamard gate $\hat{H}_{\mathrm{ad}}$ is applied $\hat{H}_{\mathrm{ad}} \ket{0} = \ket{+}$. Here $\lbrace \ket{0},\ket{1}\rbrace$ are the eigenstates of the Pauli matrix $\sigma_z$ with eigenvalues $\pm 1$, while $\lbrace \ket{+},\ket{-}\rbrace$ are the corresponding eigenstates of $\sigma_x$ with $\ket{\pm} = \left(\ket{0}\pm\ket{1}\right)/\sqrt{2}$. 
    \item The combined state of the system factorizes into $ \hat{\rho}^{\tau_0} = \hat{\rho}^{\tau_0}_{\phi} \,\otimes\, \ket{+}\bra{+}$, which undergoes a controlled unitary evolution
    \begin{equation}
        \hat{\rho}_{\mu}
        = \hat{G}_{\mu}\,\hat{\rho}^{\tau_0}\,\hat{G}_{\mu}^{\dagger},
    \end{equation}
    where the controlled–unitary operator is given by
    \begin{align}
        \hat{G}_{\mu}
        \;\equiv\; &
        \hat{U}_Ie^{-i\mu\hat{H}_{\phi}(\tau_0)}\otimes\ket{0}\bra{0}
         \\ & \;+\; \nonumber
        e^{-i\mu\hat{H}_{\phi}(\tau_1)}\,\hat{U}_I\otimes\ket{1}\bra{1}. \label{eq:G} 
    \end{align}
    $\hat{U}_I$ is given by Eq.~\eqref{eq:U_I}, and $\hat{H}_{\phi}(\tau_0) = \hat{H}_{\phi}(\tau_1) = \hat{H}_{\phi}$.
    
    \item Finally, a second Hadamard gate is applied to the UDW particle detector, resulting in the state
    \begin{equation}
        \hat{\rho}^{{\tau}}_{\mu}
        = \bigl(\hat{\mathbb{I}}\otimes\hat{H}_{\mathrm{ad}}\bigr)\,
          \hat{\rho}_{\mu}\,
          \bigl(\hat{\mathbb{I}}\otimes\hat{H}_{\mathrm{ad}}\bigr)^{\!\dagger}\,.
        \label{eq:finalstate}
    \end{equation}
\end{enumerate}

At the end of this procedure, the particle detector reduced state takes the form 
\begin{equation}
    \hat{\rho}_{UDW} = \frac{1}{2}\left(\hat{\mathbb{I}} + \mathbb{R}\text{e}(\tilde{\mathcal{P}}(\mu)) \hat{\sigma}_3 + \hat{\mathbb{I}}\text{m}(\tilde{\mathcal{P}}(\mu)) \hat{\sigma}_2 \right),
\end{equation} 
with
\begin{align}
\tilde{\mathcal{P}}(\mu) & = \Tr\left(\hat{U}^{\dagger}_I e^{i\mu\hat{H}_{\phi}} \hat{U}_I e^{-i\mu\hat{H}_{\phi}} \rho_{\phi}\right) \nonumber \\
& = \langle\hat{U}^{\dagger}_I e^{i\mu\hat{H}_{\phi}} \hat{U}_I e^{-i\mu\hat{H}_{\phi}} \rangle_{\phi} \label{eq:Pmu}
\end{align}
being the characteristic function for the field $\hat{\phi}(\mathsf{x})$ in the state $\hat{\rho}_{\phi}$. Since the characteristic function is well-defined independently of the TPM scheme, the work distribution in the Ramsey scheme can be defined as the inverse Fourier transform of $\tilde{\mathcal{P}}(\mu)$~\cite{Ortega}
\begin{align}
    \mathcal{P}(W) \equiv \frac{1}{2\pi} \int_{-\infty}^\infty \tilde{\mathcal{P}}(\mu) e^{-i W \mu}\dd\mu. \label{eq:RSwork}
\end{align}
The first point to note is that the work distribution obtained through the Ramsey scheme is a quasiprobability distribution, as discussed in Ref.~\cite{Bonfill}. In addition, the authors in Refs.~\cite{Ortega, Bonfill} demonstrated that such work distributions satisfy the Crooks detailed fluctuation theorem~\cite{Crooks1999} for KMS states, as well as the Jarzynski inequality~\cite{Jarzynski1997}, in flat spacetimes.

Hence, we show that, by constructing smeared UDW particle detectors in static curved spacetimes using Fermi normal coordinates, it is possible to extend the Ramsey protocol to obtain the characteristic function of a quantum field and subsequently define the work distribution.

%%%%%%%%%%%%%%%%%%%%%%%%%%%%%%%%%%%%%%%%%%%%%%%%%%%%%%%%%%%%
%%%%%%%%%%%%%%%%%%%%%%%%%%%%%%%%%%%%%%%%%%%%%%%%%%%%%%%%%%%%
\section{Non-perturbative characteristic function in static curved spacetimes}
\label{sec:KMSwork}

In this section, we explicitly obtain the characteristic function in a non-perturbative manner for a quantum field in a static curved spacetime. Furthermore, it is worth noting that the derivation does not rely on the field decomposition given in Eq.~\eqref{eq:phidec}.

Let us start by noting that, from Eq.~\eqref{eq:Hint}, we can rewrite Eq.~\eqref{eq:U_I} as
\begin{align}
    \hat{U}_I & = \hat{\mathcal{T}} e^{-i \lambda  \int_{\mathcal{M}} \dd^4 \mathsf{x} \sqrt{-g}f(\mathsf{x}) \hat{\phi}(\mathsf{x)}} \nonumber \\
    & = \hat{\mathcal{T}} e^{-i \lambda \hat{\phi}(f)}, \label{eq:U_I1}
\end{align}
where $f(\mathsf{x}) =  f(\mathsf{x}(\tau, \vX)) = \chi(\tau)\psi(\vX)$. By using the Magnus expansion~\cite{Blanes09, Landulfo}, we can cast Eq.~\eqref{eq:U_I1} as $\hat{U}_I = e^{\hat{\Theta}}$ where $\hat{\Theta} = \sum_{n = 1}^{\infty} \hat{\Theta}_n$. The non-zero terms of this expansion are given
\begin{align}
    & \hat{\Theta}_1 = -i \int_{-\infty}^{\infty} \dd \tau \hat{H}_I(\tau)  = - i\lambda \hat{\phi}(f),\\
    & \hat{\Theta}_2 = -\frac{1}{2} \int_{-\infty}^{\infty} \dd \tau \int_{-\infty}^{\tau} \dd \tau' [\hat{H}_I(\tau) , \hat{H}_I(\tau') ] = i \theta \hat{\mathbb{I}},
\end{align}
with
\begin{align}
    \theta = -\frac{1}{2} \lambda^2  \int_{-\infty}^{\infty} \dd \tau \int_{-\infty}^{\tau} \dd \tau' \chi(\tau) \chi(\tau') \Delta(\tau, \tau'),
\end{align}
where
\begin{small}
\begin{align}
    \Delta(\tau, \tau') = \integral_{\Sigma_\tau} \dd^3 \vX \sqrt{-g} \integral_{\Sigma_{\tau'}} \dd^3 \vX' \sqrt{-g'}  \psi(\vX)\Delta(\mathsf{x},\mathsf{x}') \psi(\vX'),
\end{align}    
\end{small}
and $\Delta(\mathsf{x},\mathsf{x}') = \Delta\left(\mathsf{x}(\tau,\vX),\mathsf{x}'(\tau', \vX')\right)$ is the unsmeared version of Eq.~\eqref{eq:Delta}, i.e, $[\hat{\phi}(\mathsf{x}), \hat{\phi}(\mathsf{x}')] = -i \Delta(\mathsf{x}, \mathsf{x}') \hat{\mathbb{I}}$. In addition, it is easy to see that $\hat{\Theta}_n = 0$ for $n \ge 3$. Hence, Eq.~\eqref{eq:U_I1} can be written as
\begin{align}
 \hat{U}_I = e^{i \theta} e^{-i \lambda \hat{\phi}(f)}.
\end{align}

In turn, this allows us to write the characteristic function for the field $\hat{\phi}(\mathsf{x})$ as
\begin{align}
    \tilde{\mathcal{P}}(\mu) = \langle e^{i\lambda\hat{\phi}(f)} e^{i\mu\hat{H}_{\phi}} e^{-i\lambda\hat{\phi}(f)} e^{-i\mu\hat{H}_{\phi}} \rangle_{\phi}.
\end{align}
By noting that 
\begin{align}
   e^{i\mu\hat{H}_{\phi}} e^{-i\lambda\hat{\phi}(f)} e^{-i\mu\hat{H}_{\phi}} & = e^{-i \lambda  \int_{\mathcal{M}} \dd^4 \mathsf{x} \sqrt{-g}f(\mathsf{x}(\tau, \vX)) \hat{\phi}(\mathsf{x}(\tau + \mu, \vX))} \nonumber \\
   & = e^{-i \lambda  \int_{\mathcal{M}} \dd^4 \mathsf{x} \sqrt{-g}f(\mathsf{x}(\tau-\mu, \vX)) \hat{\phi}(\mathsf{x}(\tau, \vX))} \nonumber\\
   & = e^{-i\lambda\hat{\phi}(g)}, 
\end{align}
where $g(\mathsf{x}(\tau, \vX)) = f(\mathsf{x}(\tau-\mu, \vX)) = \chi(\tau- \mu) \psi (\vX)$, we have
\begin{align}
     \tilde{\mathcal{P}}(\mu) = \langle e^{i\lambda\hat{\phi}(f)} e^{-i\lambda\hat{\phi}(g)} \rangle_{\phi}.
\end{align}
Finally, by making use of the Zassenhaus formula $e^{\mathfrak{a} + \mathfrak{b}}~=~e^{\mathfrak{a}}e^{\mathfrak{b}}e^{-\frac{1}{2}[\mathfrak{a},\mathfrak{b}]}$, we arrive at
\begin{align}
 \tilde{\mathcal{P}}(\mu) = e^{-\frac{i}{2}\lambda^2 \Delta(f,g)}  \langle e^{i\lambda\left(\hat{\phi}(f)-\hat{\phi}(g)\right)}  \rangle_{\phi},\label{eq:Pmu1}
\end{align}
with $ \Delta(f, g) \hat{\mathbb{I}} = i[\hat{\phi}(f), \hat{\phi}(g)] $ being defined by Eq.~\eqref{eq:Delta}. It is worth noting that Eq.~\eqref{eq:Pmu1} does not rely on the field decomposition given in Eq.~\eqref{eq:phidec}, and so far we have not assumed that the field state is a thermal KMS state. Hence, Eq.~\eqref{eq:Pmu1} is quite general, holding for a quantum field in an arbitrary state in a static curved spacetime.

\subsection{Characteristic function for thermal KMS states}
In this section, we apply the expression obtained in Eq.~\eqref{eq:Pmu1} for a thermal KMS state, i.e., $\hat{\rho}_{\phi} = \hat{\rho}_{\beta}$ and discuss some of its implications. %It is worth mentioning that characteristic function for a KMS state obtained in this section does not rely on the form of the thermal Wightman function given in Eq.~\eqref{eq:thermalW}.

Let us start by noticing that, since a KMS state is a quasifree state~\cite{Kay1991, Wald1994}, it follows that
\begin{align}
    \langle e^{i\lambda\left(\hat{\phi}(f)-\hat{\phi}(g)\right)}  \rangle_{\beta} = e^{-\frac{\lambda^2}{2}\langle(\hat{\phi}(f)-\hat{\phi}(g))^2\rangle_{\beta}}. \label{eq:aux}
\end{align}

Moreover, since the thermal Wightman function is stationary and $g(\mathsf{x})$ is related to $f(\mathsf{x})$ through a time translation, we have $\langle(\hat{\phi}(g))^2\rangle_{\beta} = \langle(\hat{\phi}(f))^2\rangle_{\beta}$ and
\begin{align}
    \langle(\hat{\phi}(f)-\hat{\phi}(g))^2\rangle_{\beta} = &  2\left(\langle(\hat{\phi}(f))^2\rangle_{\beta}  - \langle\hat{\phi}(f)\hat{\phi}(g)\rangle_{\beta}\right) \nonumber \\ & + i \Delta(f,g).
\end{align}
Together with Eqs.~\eqref{eq:Pmu1} and~\eqref{eq:aux}, this gives us the final form of the characteristic function of a KMS state, namely
\begin{align}
  \tilde{\mathcal{P}}(\mu) =    e^{\lambda^2 \left(\langle\hat{\phi}(f)\hat{\phi}(g)\rangle_{\beta} - \langle(\hat{\phi}(f))^2\rangle_{\beta}\right)} , \label{eq:PmuKMS}
\end{align}
which coincides with the expression obtained in Ref.~\cite{Bonfill} for the characteristic function of a KMS state of a quantum field in a flat spacetime. In contrast, here we derive the same expression for a static curved spacetime.

In terms of the thermal Wightman functions, Eq.~\eqref{eq:PmuKMS} can be written as
\begin{align}
 \tilde{\mathcal{P}}(\mu) = &  \exp\{\lambda^2  \int_{-\infty}^{\infty} \dd \tau'  \chi(\tau') \int_{-\infty}^{\infty} \dd \tau \chi(\tau) \nonumber  \\ & \times \integral_{\Sigma_{\tau'}} \dd^3 \vX' \sqrt{-g'}  \psi(\vX') \integral_{\Sigma_\tau} \dd^3 \vX \sqrt{-g} \psi(\vX)  \nonumber \\ & \times \left(\mathcal{W}_{\beta}(\tau', \tau+ \mu) - \mathcal{W}_{\beta}(\tau', \tau) \right)\}, \label{eq:PmuKMS1}
\end{align}
where $\mathcal{W}_{\beta}(\tau, \tau')$ is the pullback of the Wightman function along the worldline $\gamma(\tau)$. 

The expression~\eqref{eq:PmuKMS1} allows us to recover the detailed Crooks theorem~\cite{Crooks1999} and the Jarzynski equality~\cite{Jarzynski1997} solely from the properties of the thermal Wightman functions, i.e., from the KMS condition. Indeed, since $\mathcal{W}_{\beta}(\tau'~,~\tau~+~\mu)~=~\mathcal{W}_{\beta}(-\Delta\tau-\mu)$ and $\mathcal{W}_{\beta}(\tau'~,~\tau)~=~\mathcal{W}_{\beta}(- \Delta\tau)$, and noting that $\mathcal{W}_{\beta}(-~\Delta\tau~+~\mu~-~i~\beta)~=~\mathcal{W}_{\beta}(\Delta\tau~-~\mu)$, it is straightforward to see that
\begin{equation}
    \tilde{\mathcal{P}}(-\mu + i \beta) = \tilde{\mathcal{P}}(\mu) .
\end{equation}
From the definition of the Ramsey scheme work distribution given by Eq.~\eqref{eq:RSwork}, it follows that $\mathcal{P}(W)/\mathcal{P}_{\text{rev}}(-W)~=~e^{\beta W}$, being a particular case of Crooks theorem in which $\Delta F =0$, since the protocol described in Sec.~\ref{sec:ramsey} assumes that $\hat{H}_{\phi}(\tau_0) = \hat{H}_{\phi}(\tau_1) = \hat{H}_{\phi}$. This fluctuation relation establishes that the probability distribution $\mathcal{P}(W)$ for work performed during the forward process and the probability distribution $\mathcal{P}_{\text{rev}}(-W)$ for the reverse process are related by an exponential factor that depends on the positive work $W$, demonstrating that positive work values are exponentially more probable than their negative counterparts in the reverse process.
Moreover, this can be interpreted as a detailed balance condition for the work distribution, rather than for the transition rates or for the Fourier transform of the Wightman function~\cite{Fewster, Aubry, Perche2021}.

Direct integration of Crooks' theorem yields the Jarzynski equality, $\langle e^{-\beta W}\rangle = 1$, for the $\Delta F = 0$ case. An alternative derivation follows directly from the characteristic function, since $\tilde{\mathcal{P}}(i\beta) = 1$, which can be verified by explicit evaluation of Eq.~\eqref{eq:PmuKMS1} at $\mu = i\beta$. The apparent violation of the second law suggested by transient negative work values ($W < 0$) in individual realizations is precisely compensated by rare events with large positive work contributions, thereby ensuring $\langle W \rangle \geq 0$~\cite{Jarzynski11}. 

%%%%%%%%%%%%%%%%%%%%%%%%%%%%%%%%%%%%%%%%%%%%%%%%%%%%%%%%%
%%%%%%%%%%%%%%%%%%%%%%%%%%%%%%%%%%%%%%%%%%%%%%%%%%%%%%%%%
\subsection{Pointlike detector}
In this section, we consider the particular case of a pointlike detector that follows the same worldline $\gamma(\tau)$ as our observer $\mathcal{O}$. The pointlike nature of the detector allows us to exploit the decomposition of the field given in Eq.~\eqref{eq:phidec} in coordinates adapted to the spacetime symmetry and in which the field is quantized, providing a direct relation with the Fermi normal coordinates adapted to the detector's frame. This allows us to obtain a useful expression for Eq.~\eqref{eq:PmuKMS1}. 

In this case, considering a pointlike detector amounts to choosing the smearing function to be~\cite{Poisson11}
\begin{align}
    \psi(\mathsf{x}(\tau, \vX)) = \frac{1}{\sqrt{-g}}\int \delta^4 \left((\mathsf{x(\tau)- \gamma(\tau)})\right)\dd\tau,\label{eq:pointdet}
\end{align}
where the integral is taken over the proper time $\tau$ of the worldline $\gamma(\tau)$. As discussed in Sec.~\ref{sec:fermi}, the worldline $\gamma(\tau)$ can be expressed as $\gamma(\tau) = (N^{-1}(\vx_0)(\tau_0 + \tau), \vx_0)$. In terms of the Fermi coordinates, we have $\vX_0 = \vX_0(\vx_0)$ and $\tau =N(\vx_0) t$.

Moreover, the field decomposition along the worldline $\gamma(\tau)$ has the same form as Eq.~\eqref{eq:phidec} with $\hat{\phi}(\mathsf{x}(\tau, \vX_0))~\equiv~\hat{\phi}(\tau, \vx_0)$ and with the frequencies $\omega_j$ shifted to $\Omega_j = N^{-1}(\vx_0) \omega_j$. In this case, with the use of Eq.~\eqref{eq:phidec}, the thermal Wightman function at inverse temperature $1/\beta$ in a static spacetime can be written as~\cite{Aubry}
\begin{align}
 \mathcal{W}_{\beta}(\Delta \tau) = \int\frac{\dd \xi (j)}{2 \Omega_j}   F_j(\vx_0)F^*_j(\vx'_0) \frac{\cos\left[\Omega_j(\Delta \tau + i\beta/2)\right]}{\sinh(\beta \Omega_j /2)}.\label{eq:thermalW}
\end{align}

By using Eqs.~\eqref{eq:pointdet} and~\eqref{eq:thermalW} into Eq.~\eqref{eq:PmuKMS1}, along with the fact that $\vx_0 = \vx'_0$, a straightforward calculation allows us to obtain
\begin{align}
 \tilde{\mathcal{P}}(\mu) = &  \exp\{\lambda^2  \int \frac{\dd \xi(j)}{2\Omega_j} \frac{|\tilde{\chi}(\Omega_j)|^2}{\sinh(\beta \Omega_j/2)}|F_j(\vx_0)|^2\nonumber \\ 
 & \times \left(\cos [\Omega_j(\mu - i\beta/2)] - \cosh(\beta \Omega_j/2) \right)\},\label{eq:PmuKMS2}    
\end{align}
where $\tilde{\chi}(\Omega_j)$ denotes the Fourier transform of $\chi(\tau)$ with respect to $\tau$.

Following Ref.~\cite{Ortega} and using Eq.~\eqref{eq:PmuKMS2}, we are able to compute the first two moments of the distribution $\mathcal{P}(W)$, i.e.,
 \begin{align}
\langle W\rangle_{\beta} &= i^{-1}\left(\dfrac{\dd\tilde{\mathcal{P}}}{\dd\mu}\right)_{\mu = 0} \label{eq:moment1} \\ &= \dfrac{\lambda^2}{2}\integral \dd\xi(j) \left|\tilde{\chi}(\Omega_j)\right|^2 |F_j(\vx_0)|^2, \nonumber \\
\langle W^2\rangle_{\beta} &= -\left(\dfrac{\dd^2\tilde{\mathcal{P}}}{\dd\mu^2}\right)_{\mu = 0}\label{eq:moment2} \\
&= \dfrac{\lambda^2}{2}\integral \dd\xi(j) \Omega_j\left|\tilde{\chi}(\Omega_j)\right|^2\coth\left(\dfrac{\beta\Omega_j}{2}\right) |F_j(\vx_0)|^2, \nonumber
\end{align}
where $|F_j(\vx_0)|^2$ denotes the squared modulus of mode $j$ of the field, evaluated at the worldline $\gamma(\tau)$, and is therefore a constant. From Eq.~\eqref{eq:moment1}, we can see that $\langle W\rangle_{\beta} \ge 0$, as expected, and which provides information about the averaged work required to implement the localized unitary on a quantum field. Moreover, Eq.~\eqref{eq:moment2} is calculated up to order $O(\lambda^2)$ and explicitly depends on the field spectrum.

In addition, the work variance $\sigma^2_{\beta} = \langle W^2\rangle_{\beta} - \langle W\rangle_{\beta}^2$ becomes
\begin{align}
	\sigma^2_{\beta} & = \dfrac{\lambda^2}{2}\integral \dd\xi(j) \Omega_j\left|\tilde{\chi}(\Omega_j)\right|^2\coth\left(\dfrac{\beta\Omega_j}{2}\right) |F_j(\vx_0)|^2 \nonumber \\ 
    & \ + O(\lambda^4) \ ,
	\label{eq:variance}
\end{align}
which monotonically increases with $1/\beta$, as expected from thermal fluctuations. Moreover, in order to analyze the connection between the average work and the work variance, let us consider the high temperature limit $\beta \Omega_j \ll 1$. In this limit, we have $\coth\left(\beta\Omega_j/2\right) \approx 2/\beta\Omega_j$ and the average work becomes proportional to the work variance
\begin{align}
    \langle W\rangle_{\beta} \approx \frac{1}{2}\beta \sigma^2_{\beta}
\end{align}
thus recovering the standard fluctuation-dissipation relation within linear response theory~\cite{Jarzynski11}.

%%%%%%%%%%%%%%%%%%%%%%%%%%%%%%%%%%%%%%%%%%%%%%%%%%%%%%%%%%%%%
%%%%%%%%%%%%%%%%%%%%%%%%%%%%%%%%%%%%%%%%%%%%%%%%%%%%%%%%%%%%%

%%%%%%%%%%%%%%%%%%%%%%%%%%%%%%%%%%%%%%%%%%%%%%%%%%%%%%%%%%%%%
%%%%%%%%%%%%%%%%%%%%%%%%%%%%%%%%%%%%%%%%%%%%%%%%%%%%%%%%%%%%%
\section{Conclusions}
\label{conclusions}

In this work, we extended the procedure established in Ref.~\cite{Ortega} to define the Ramsey scheme work distribution for a quantum field in a static curved spacetime. This setup overcomes the causality violations inherent in the conventional two-time measurement protocol and ensures operational consistency in relativistic quantum field theory. 

By analyzing the non-perturbative expression for the characteristic function of the quantum field, we have shown that fluctuation theorems emerge from the underlying Kubo–Martin–Schwinger condition satisfied by the initial thermal state of the field. Moreover, we discussed the example of a pointlike detector interacting with the quantum field, which allows us to obtain a relatively simple expression for the characteristic function and to determine the first two moments of the work distribution, as well as the work variance, which in the high-temperature limit reproduces the standard fluctuation-dissipation relation.

The emergence of the fluctuation theorems in this context demonstrates that thermodynamic irreversibility persists at the level of quantum field interactions in curved spacetimes, and that its signature can be captured through local interactions with UDW particle detectors. The forward and reverse processes, related by time-reversal symmetry, obey a detailed balance relation that reflects the KMS structure of the initial state. Importantly, this result holds despite the infinite number of degrees of freedom in the quantum field, affirming the robustness of fluctuation relations even in quantum field theoretic settings. 

Several promising directions emerge from this framework. One natural extension is to consider more general spacetimes, where the absence of a global timelike Killing vector complicates the definition of equilibrium states. Up to Eq.~\eqref{eq:Pmu1}, we have not relied on the field decomposition that is valid only for static spacetimes, nor on the KMS condition; therefore, Eq.~\eqref{eq:Pmu1} remains valid for general globally hyperbolic curved spacetimes. However, starting from Eq.~\eqref{eq:Pmu1}, further investigation is required, possibly making use of local KMS equilibrium states~\cite{Solveen12,Gransee17}, which will be addressed in a future work. Regarding static spacetimes, it would also be interesting to consider families of observers different from those following the flow of the timelike Killing vector field. This could have implications for entropy production as perceived by different observers~\cite{Wald01,Marolf04,Basso2025} and could lead to a breakdown of the detailed balance condition.

Another important consideration in this context is the inclusion of the entropy production in the detector itself due to the coupling to the gravitational field~\cite{Basso2023,Basso2025}. Moreover, entropy is generated in the field if it is subject to dynamical boundary conditions~\cite{Oliveira2024}. Such an entropy, which is deeply linked to quantum coherence and entanglement, is expected to be present also in the case of a time-dependent metric. The contribution of the quantum degrees of freedom of the (weak) gravitational field~\cite{Moreira2025} may also be important within this framework. Therefore, the investigation of a fluctuation relation encompassing all of these contributions may provide deeper insights into the thermodynamics of relativistic quantum fields.   

Ultimately, as relativistic quantum information theory continues to evolve, the operational formulation, such as the one presented here, will play a central role in bridging the conceptual gap between thermodynamics, quantum field theory, and general relativity.

%%%%%%%%%%%%%%%%%%%%%%%%%%%%%%%%%%%%%%%%%%%%%%%%%%%%%%%%%%%%%%%
%%%%%%%%%%%%%%%%%%%%%%%%%%%%%%%%%%%%%%%%%%%%%%%%%%%%%%%%%%%%%%%
\begin{acknowledgments}
This work was supported by the Coordination for the Improvement of Higher Education Personnel (CAPES), S\~{a}o Paulo Research Foundation (FAPESP), Grant No.~2025/07325-0, by the National Institute for the Science and Technology of Quantum Information (INCT-IQ), Grant No.~465469/2014-0, and by the National Council for Scientific and Technological Development (CNPq), Grants No.~300083/2025-4 and No~308065/2022-0.
\end{acknowledgments}


\begin{thebibliography}{65}%
\makeatletter
\providecommand \@ifxundefined [1]{%
 \@ifx{#1\undefined}
}%
\providecommand \@ifnum [1]{%
 \ifnum #1\expandafter \@firstoftwo
 \else \expandafter \@secondoftwo
 \fi
}%
\providecommand \@ifx [1]{%
 \ifx #1\expandafter \@firstoftwo
 \else \expandafter \@secondoftwo
 \fi
}%
\providecommand \natexlab [1]{#1}%
\providecommand \enquote  [1]{``#1''}%
\providecommand \bibnamefont  [1]{#1}%
\providecommand \bibfnamefont [1]{#1}%
\providecommand \citenamefont [1]{#1}%
\providecommand \href@noop [0]{\@secondoftwo}%
\providecommand \href [0]{\begingroup \@sanitize@url \@href}%
\providecommand \@href[1]{\@@startlink{#1}\@@href}%
\providecommand \@@href[1]{\endgroup#1\@@endlink}%
\providecommand \@sanitize@url [0]{\catcode `\\12\catcode `\$12\catcode
  `\&12\catcode `\#12\catcode `\^12\catcode `\_12\catcode `\%12\relax}%
\providecommand \@@startlink[1]{}%
\providecommand \@@endlink[0]{}%
\providecommand \url  [0]{\begingroup\@sanitize@url \@url }%
\providecommand \@url [1]{\endgroup\@href {#1}{\urlprefix }}%
\providecommand \urlprefix  [0]{URL }%
\providecommand \Eprint [0]{\href }%
\providecommand \doibase [0]{https://doi.org/}%
\providecommand \selectlanguage [0]{\@gobble}%
\providecommand \bibinfo  [0]{\@secondoftwo}%
\providecommand \bibfield  [0]{\@secondoftwo}%
\providecommand \translation [1]{[#1]}%
\providecommand \BibitemOpen [0]{}%
\providecommand \bibitemStop [0]{}%
\providecommand \bibitemNoStop [0]{.\EOS\space}%
\providecommand \EOS [0]{\spacefactor3000\relax}%
\providecommand \BibitemShut  [1]{\csname bibitem#1\endcsname}%
\let\auto@bib@innerbib\@empty
%</preamble>
\bibitem [{\citenamefont {Hawking}(1975)}]{Hawking1975}%
  \BibitemOpen
  \bibfield  {author} {\bibinfo {author} {\bibfnamefont {S.~W.}\ \bibnamefont
  {Hawking}},\ }\bibfield  {title} {\bibinfo {title} {Particle creation by
  black holes},\ }\href {https://doi.org/10.1007/BF02345020} {\bibfield
  {journal} {\bibinfo  {journal} {Commun. Math. Phys.}\ }\textbf {\bibinfo
  {volume} {43}},\ \bibinfo {pages} {199} (\bibinfo {year} {1975})}\BibitemShut
  {NoStop}%
\bibitem [{\citenamefont {Bekenstein}(1973)}]{Bekenstein1973}%
  \BibitemOpen
  \bibfield  {author} {\bibinfo {author} {\bibfnamefont {J.~D.}\ \bibnamefont
  {Bekenstein}},\ }\bibfield  {title} {\bibinfo {title} {Black holes and
  entropy},\ }\href {https://doi.org/10.1103/PhysRevD.7.2333} {\bibfield
  {journal} {\bibinfo  {journal} {Phys. Rev. D}\ }\textbf {\bibinfo {volume}
  {7}},\ \bibinfo {pages} {2333} (\bibinfo {year} {1973})}\BibitemShut
  {NoStop}%
\bibitem [{\citenamefont {Birrell}\ and\ \citenamefont
  {Davies}(1982)}]{Birrell1982}%
  \BibitemOpen
  \bibfield  {author} {\bibinfo {author} {\bibfnamefont {N.~D.}\ \bibnamefont
  {Birrell}}\ and\ \bibinfo {author} {\bibfnamefont {P.~C.~W.}\ \bibnamefont
  {Davies}},\ }\href {https://doi.org/10.1017/CBO9780511622632} {\emph
  {\bibinfo {title} {Quantum Fields in Curved Space}}}\ (\bibinfo  {publisher}
  {Cambridge University Press},\ \bibinfo {address} {Cambridge},\ \bibinfo
  {year} {1982})\BibitemShut {NoStop}%
\bibitem [{\citenamefont {Wald}(1994)}]{Wald1994}%
  \BibitemOpen
  \bibfield  {author} {\bibinfo {author} {\bibfnamefont {R.~M.}\ \bibnamefont
  {Wald}},\ }\href@noop {} {\emph {\bibinfo {title} {Quantum Field Theory in
  Curved Spacetime and Black Hole Thermodynamics}}}\ (\bibinfo  {publisher}
  {University of Chicago Press},\ \bibinfo {address} {Chicago},\ \bibinfo
  {year} {1994})\BibitemShut {NoStop}%
\bibitem [{\citenamefont {Crooks}(1999)}]{Crooks1999}%
  \BibitemOpen
  \bibfield  {author} {\bibinfo {author} {\bibfnamefont {G.~E.}\ \bibnamefont
  {Crooks}},\ }\bibfield  {title} {\bibinfo {title} {Entropy production
  fluctuation theorem and the nonequilibrium work relation for free energy
  differences},\ }\href {https://doi.org/10.1103/PhysRevE.60.2721} {\bibfield
  {journal} {\bibinfo  {journal} {Phys. Rev. E}\ }\textbf {\bibinfo {volume}
  {60}},\ \bibinfo {pages} {2721} (\bibinfo {year} {1999})}\BibitemShut
  {NoStop}%
\bibitem [{\citenamefont {Jarzynski}(1997)}]{Jarzynski1997}%
  \BibitemOpen
  \bibfield  {author} {\bibinfo {author} {\bibfnamefont {C.}~\bibnamefont
  {Jarzynski}},\ }\bibfield  {title} {\bibinfo {title} {Nonequilibrium equality
  for free energy differences},\ }\href
  {https://doi.org/10.1103/PhysRevLett.78.2690} {\bibfield  {journal} {\bibinfo
   {journal} {Phys. Rev. Lett.}\ }\textbf {\bibinfo {volume} {78}},\ \bibinfo
  {pages} {2690} (\bibinfo {year} {1997})}\BibitemShut {NoStop}%
\bibitem [{\citenamefont {Talkner}\ \emph {et~al.}(2007)\citenamefont
  {Talkner}, \citenamefont {Lutz},\ and\ \citenamefont
  {H\"anggi}}]{Talkner2007}%
  \BibitemOpen
  \bibfield  {author} {\bibinfo {author} {\bibfnamefont {P.}~\bibnamefont
  {Talkner}}, \bibinfo {author} {\bibfnamefont {E.}~\bibnamefont {Lutz}},\ and\
  \bibinfo {author} {\bibfnamefont {P.}~\bibnamefont {H\"anggi}},\ }\bibfield
  {title} {\bibinfo {title} {Fluctuation theorems: Work is not an observable},\
  }\href {https://doi.org/10.1103/PhysRevE.75.050102} {\bibfield  {journal}
  {\bibinfo  {journal} {Phys. Rev. E}\ }\textbf {\bibinfo {volume} {75}},\
  \bibinfo {pages} {050102(R)} (\bibinfo {year} {2007})}\BibitemShut {NoStop}%
\bibitem [{\citenamefont {Dorner}\ \emph {et~al.}(2013)\citenamefont {Dorner},
  \citenamefont {Clark}, \citenamefont {Heaney}, \citenamefont {Fazio},
  \citenamefont {Calarco},\ and\ \citenamefont {Jaksch}}]{Dorner2013}%
  \BibitemOpen
  \bibfield  {author} {\bibinfo {author} {\bibfnamefont {R.}~\bibnamefont
  {Dorner}}, \bibinfo {author} {\bibfnamefont {S.~R.}\ \bibnamefont {Clark}},
  \bibinfo {author} {\bibfnamefont {L.}~\bibnamefont {Heaney}}, \bibinfo
  {author} {\bibfnamefont {R.}~\bibnamefont {Fazio}}, \bibinfo {author}
  {\bibfnamefont {T.}~\bibnamefont {Calarco}},\ and\ \bibinfo {author}
  {\bibfnamefont {D.}~\bibnamefont {Jaksch}},\ }\bibfield  {title} {\bibinfo
  {title} {Extracting quantum work statistics and fluctuation theorems by
  single-qubit interferometry},\ }\href
  {https://doi.org/10.1103/PhysRevLett.110.230601} {\bibfield  {journal}
  {\bibinfo  {journal} {Phys. Rev. Lett.}\ }\textbf {\bibinfo {volume} {110}},\
  \bibinfo {pages} {230601} (\bibinfo {year} {2013})}\BibitemShut {NoStop}%
\bibitem [{\citenamefont {Mazzola}\ \emph {et~al.}(2013)\citenamefont
  {Mazzola}, \citenamefont {Chiara},\ and\ \citenamefont
  {Paternostro}}]{Mazzola2013}%
  \BibitemOpen
  \bibfield  {author} {\bibinfo {author} {\bibfnamefont {L.}~\bibnamefont
  {Mazzola}}, \bibinfo {author} {\bibfnamefont {G.~D.}\ \bibnamefont
  {Chiara}},\ and\ \bibinfo {author} {\bibfnamefont {M.}~\bibnamefont
  {Paternostro}},\ }\bibfield  {title} {\bibinfo {title} {Measuring the
  characteristic function of the work distribution},\ }\href
  {https://doi.org/10.1103/PhysRevLett.110.230602} {\bibfield  {journal}
  {\bibinfo  {journal} {Phys. Rev. Lett.}\ }\textbf {\bibinfo {volume} {110}},\
  \bibinfo {pages} {230602} (\bibinfo {year} {2013})}\BibitemShut {NoStop}%
\bibitem [{\citenamefont {Campisi}\ \emph {et~al.}(2011)\citenamefont
  {Campisi}, \citenamefont {Talkner},\ and\ \citenamefont
  {H\"anggi}}]{Campisi2011}%
  \BibitemOpen
  \bibfield  {author} {\bibinfo {author} {\bibfnamefont {M.}~\bibnamefont
  {Campisi}}, \bibinfo {author} {\bibfnamefont {P.}~\bibnamefont {Talkner}},\
  and\ \bibinfo {author} {\bibfnamefont {P.}~\bibnamefont {H\"anggi}},\
  }\bibfield  {title} {\bibinfo {title} {Colloquium: Quantum fluctuation
  relations: Foundations and applications},\ }\href
  {https://doi.org/10.1103/RevModPhys.83.771} {\bibfield  {journal} {\bibinfo
  {journal} {Rev. Mod. Phys.}\ }\textbf {\bibinfo {volume} {83}},\ \bibinfo
  {pages} {771} (\bibinfo {year} {2011})}\BibitemShut {NoStop}%
\bibitem [{\citenamefont {Esposito}\ \emph {et~al.}(2009)\citenamefont
  {Esposito}, \citenamefont {Harbola},\ and\ \citenamefont
  {Mukamel}}]{Esposito2009}%
  \BibitemOpen
  \bibfield  {author} {\bibinfo {author} {\bibfnamefont {M.}~\bibnamefont
  {Esposito}}, \bibinfo {author} {\bibfnamefont {U.}~\bibnamefont {Harbola}},\
  and\ \bibinfo {author} {\bibfnamefont {S.}~\bibnamefont {Mukamel}},\
  }\bibfield  {title} {\bibinfo {title} {Nonequilibrium fluctuations,
  fluctuation theorems, and counting statistics in quantum systems},\ }\href
  {https://doi.org/10.1103/RevModPhys.81.1665} {\bibfield  {journal} {\bibinfo
  {journal} {Rev. Mod. Phys.}\ }\textbf {\bibinfo {volume} {81}},\ \bibinfo
  {pages} {1665} (\bibinfo {year} {2009})}\BibitemShut {NoStop}%
\bibitem [{\citenamefont {Kafri}\ and\ \citenamefont
  {Deffner}(2012)}]{Kafri2012}%
  \BibitemOpen
  \bibfield  {author} {\bibinfo {author} {\bibfnamefont {D.}~\bibnamefont
  {Kafri}}\ and\ \bibinfo {author} {\bibfnamefont {S.}~\bibnamefont
  {Deffner}},\ }\bibfield  {title} {\bibinfo {title} {Holevo's bound from a
  general quantum fluctuation theorem},\ }\href
  {https://doi.org/10.1103/PhysRevA.86.044302} {\bibfield  {journal} {\bibinfo
  {journal} {Phys. Rev. A}\ }\textbf {\bibinfo {volume} {86}},\ \bibinfo
  {pages} {044302} (\bibinfo {year} {2012})}\BibitemShut {NoStop}%
\bibitem [{\citenamefont {Rastegin}\ and\ \citenamefont
  {Życzkowski}(2014)}]{Rastegin2014}%
  \BibitemOpen
  \bibfield  {author} {\bibinfo {author} {\bibfnamefont {A.~E.}\ \bibnamefont
  {Rastegin}}\ and\ \bibinfo {author} {\bibfnamefont {K.}~\bibnamefont
  {Życzkowski}},\ }\bibfield  {title} {\bibinfo {title} {Jarzynski equality
  for quantum stochastic maps},\ }\href
  {https://doi.org/10.1103/PhysRevE.89.012127} {\bibfield  {journal} {\bibinfo
  {journal} {Phys. Rev. E}\ }\textbf {\bibinfo {volume} {89}},\ \bibinfo
  {pages} {012127} (\bibinfo {year} {2014})}\BibitemShut {NoStop}%
\bibitem [{\citenamefont {Sorkin}(1993)}]{Sorkin1993}%
  \BibitemOpen
  \bibfield  {author} {\bibinfo {author} {\bibfnamefont {R.~D.}\ \bibnamefont
  {Sorkin}},\ }\bibfield  {title} {\bibinfo {title} {Impossible measurements on
  quantum fields},\ }\href {https://arxiv.org/abs/gr-qc/9302018v2} {\bibfield
  {journal} {\bibinfo  {journal} {Directions in General Relativity}\ }\textbf
  {\bibinfo {volume} {2}},\ \bibinfo {pages} {293} (\bibinfo {year}
  {1993})}\BibitemShut {NoStop}%
\bibitem [{\citenamefont {Borsten}\ \emph {et~al.}(2021)\citenamefont
  {Borsten}, \citenamefont {Jubb},\ and\ \citenamefont {Kells}}]{Borsten}%
  \BibitemOpen
  \bibfield  {author} {\bibinfo {author} {\bibfnamefont {L.}~\bibnamefont
  {Borsten}}, \bibinfo {author} {\bibfnamefont {I.}~\bibnamefont {Jubb}},\ and\
  \bibinfo {author} {\bibfnamefont {G.}~\bibnamefont {Kells}},\ }\bibfield
  {title} {\bibinfo {title} {Impossible measurements revisited},\ }\href
  {https://doi.org/10.1103/PhysRevD.104.025012} {\bibfield  {journal} {\bibinfo
   {journal} {Phys. Rev. D}\ }\textbf {\bibinfo {volume} {104}},\ \bibinfo
  {pages} {025012} (\bibinfo {year} {2021})}\BibitemShut {NoStop}%
\bibitem [{\citenamefont {Bostelmann}\ \emph {et~al.}(2021)\citenamefont
  {Bostelmann}, \citenamefont {Fewster},\ and\ \citenamefont
  {Ruep}}]{Bostelmann}%
  \BibitemOpen
  \bibfield  {author} {\bibinfo {author} {\bibfnamefont {H.}~\bibnamefont
  {Bostelmann}}, \bibinfo {author} {\bibfnamefont {C.~J.}\ \bibnamefont
  {Fewster}},\ and\ \bibinfo {author} {\bibfnamefont {M.~H.}\ \bibnamefont
  {Ruep}},\ }\bibfield  {title} {\bibinfo {title} {Impossible measurements
  require impossible apparatus},\ }\href
  {https://doi.org/10.1103/PhysRevD.103.025017} {\bibfield  {journal} {\bibinfo
   {journal} {Phys. Rev. D}\ }\textbf {\bibinfo {volume} {103}},\ \bibinfo
  {pages} {025017} (\bibinfo {year} {2021})}\BibitemShut {NoStop}%
\bibitem [{\citenamefont {Anastopoulos}\ and\ \citenamefont
  {Savvidou}(2022)}]{Anastopoulos2022}%
  \BibitemOpen
  \bibfield  {author} {\bibinfo {author} {\bibfnamefont {C.}~\bibnamefont
  {Anastopoulos}}\ and\ \bibinfo {author} {\bibfnamefont {N.}~\bibnamefont
  {Savvidou}},\ }\bibfield  {title} {\bibinfo {title} {Quantum information in
  relativity: The challenge of qft measurements},\ }\bibfield  {journal}
  {\bibinfo  {journal} {Entropy}\ }\textbf {\bibinfo {volume} {24}},\ \href
  {https://doi.org/10.3390/e24010004} {10.3390/e24010004} (\bibinfo {year}
  {2022})\BibitemShut {NoStop}%
\bibitem [{\citenamefont {Unruh}(1976)}]{Unruh1976}%
  \BibitemOpen
  \bibfield  {author} {\bibinfo {author} {\bibfnamefont {W.~G.}\ \bibnamefont
  {Unruh}},\ }\bibfield  {title} {\bibinfo {title} {Notes on black-hole
  evaporation},\ }\href {https://doi.org/10.1103/PhysRevD.14.870} {\bibfield
  {journal} {\bibinfo  {journal} {Phys. Rev. D}\ }\textbf {\bibinfo {volume}
  {14}},\ \bibinfo {pages} {870} (\bibinfo {year} {1976})}\BibitemShut
  {NoStop}%
\bibitem [{\citenamefont {DeWitt}(1979)}]{DeWitt1979}%
  \BibitemOpen
  \bibfield  {author} {\bibinfo {author} {\bibfnamefont {B.~S.}\ \bibnamefont
  {DeWitt}},\ }\bibfield  {title} {\bibinfo {title} {Quantum gravity: the new
  synthesis},\ }in\ \href@noop {} {\emph {\bibinfo {booktitle} {General
  Relativity: An Einstein Centenary Survey}}},\ \bibinfo {editor} {edited by\
  \bibinfo {editor} {\bibfnamefont {S.~W.}\ \bibnamefont {Hawking}}\ and\
  \bibinfo {editor} {\bibfnamefont {W.}~\bibnamefont {Israel}}}\ (\bibinfo
  {publisher} {Cambridge University Press},\ \bibinfo {year} {1979})\ pp.\
  \bibinfo {pages} {680--745}\BibitemShut {NoStop}%
\bibitem [{\citenamefont {Takagi}(1986)}]{Takagi1986}%
  \BibitemOpen
  \bibfield  {author} {\bibinfo {author} {\bibfnamefont {S.}~\bibnamefont
  {Takagi}},\ }\bibfield  {title} {\bibinfo {title} {Vacuum noise and stress
  induced by uniform accelerator: Hawking-unruh effect in rindler manifold of
  arbitrary dimension},\ }\href {https://doi.org/10.1143/PTPS.88.1} {\bibfield
  {journal} {\bibinfo  {journal} {Progress of Theoretical Physics Supplement}\
  }\textbf {\bibinfo {volume} {88}},\ \bibinfo {pages} {1} (\bibinfo {year}
  {1986})}\BibitemShut {NoStop}%
\bibitem [{\citenamefont {Schlicht}(2004)}]{Schlicht04}%
  \BibitemOpen
  \bibfield  {author} {\bibinfo {author} {\bibfnamefont {S.}~\bibnamefont
  {Schlicht}},\ }\bibfield  {title} {\bibinfo {title} {Considerations on the
  unruh effect: causality and regularization},\ }\href
  {https://doi.org/10.1088/0264-9381/21/19/011} {\bibfield  {journal} {\bibinfo
   {journal} {Class. Quantum Grav}\ }\textbf {\bibinfo {volume} {21}},\
  \bibinfo {pages} {4647} (\bibinfo {year} {2004})}\BibitemShut {NoStop}%
\bibitem [{\citenamefont {Louko}\ and\ \citenamefont {Satz}(2006)}]{Louko06}%
  \BibitemOpen
  \bibfield  {author} {\bibinfo {author} {\bibfnamefont {J.}~\bibnamefont
  {Louko}}\ and\ \bibinfo {author} {\bibfnamefont {A.}~\bibnamefont {Satz}},\
  }\bibfield  {title} {\bibinfo {title} {How often does the unruh–dewitt
  detector click? regularization by a spatial profile},\ }\href
  {https://doi.org/10.1088/0264-9381/23/22/015} {\bibfield  {journal} {\bibinfo
   {journal} {Class. and Quantum Grav.}\ }\textbf {\bibinfo {volume} {23}},\
  \bibinfo {pages} {6321} (\bibinfo {year} {2006})}\BibitemShut {NoStop}%
\bibitem [{\citenamefont {Louko}\ and\ \citenamefont {Satz}(2008)}]{Louko08}%
  \BibitemOpen
  \bibfield  {author} {\bibinfo {author} {\bibfnamefont {J.}~\bibnamefont
  {Louko}}\ and\ \bibinfo {author} {\bibfnamefont {A.}~\bibnamefont {Satz}},\
  }\bibfield  {title} {\bibinfo {title} {Transition rate of the unruh–dewitt
  detector in curved spacetime},\ }\href
  {https://doi.org/10.1088/0264-9381/25/5/055012} {\bibfield  {journal}
  {\bibinfo  {journal} {Class. Quantum Grav.}\ }\textbf {\bibinfo {volume}
  {25}},\ \bibinfo {pages} {055012} (\bibinfo {year} {2008})}\BibitemShut
  {NoStop}%
\bibitem [{\citenamefont {Satz}(2007)}]{Satz2007}%
  \BibitemOpen
  \bibfield  {author} {\bibinfo {author} {\bibfnamefont {A.}~\bibnamefont
  {Satz}},\ }\bibfield  {title} {\bibinfo {title} {Then again, how often does
  the unruh-dewitt detector click if we switch it carefully?},\ }\href
  {https://doi.org/10.1088/0264-9381/24/7/003} {\bibfield  {journal} {\bibinfo
  {journal} {Class. Quantum Grav.}\ }\textbf {\bibinfo {volume} {24}},\
  \bibinfo {pages} {1719} (\bibinfo {year} {2007})}\BibitemShut {NoStop}%
\bibitem [{\citenamefont {de~Ram\'on}\ \emph {et~al.}(2021)\citenamefont
  {de~Ram\'on}, \citenamefont {Papageorgiou},\ and\ \citenamefont
  {Mart\'{\i}n-Mart\'{\i}nez}}]{Ramon21}%
  \BibitemOpen
  \bibfield  {author} {\bibinfo {author} {\bibfnamefont {J.}~\bibnamefont
  {de~Ram\'on}}, \bibinfo {author} {\bibfnamefont {M.}~\bibnamefont
  {Papageorgiou}},\ and\ \bibinfo {author} {\bibfnamefont {E.}~\bibnamefont
  {Mart\'{\i}n-Mart\'{\i}nez}},\ }\bibfield  {title} {\bibinfo {title}
  {Relativistic causality in particle detector models: Faster-than-light
  signaling and impossible measurements},\ }\href
  {https://doi.org/10.1103/PhysRevD.103.085002} {\bibfield  {journal} {\bibinfo
   {journal} {Phys. Rev. D}\ }\textbf {\bibinfo {volume} {103}},\ \bibinfo
  {pages} {085002} (\bibinfo {year} {2021})}\BibitemShut {NoStop}%
\bibitem [{\citenamefont {Tjoa}(2022)}]{Tjoa22}%
  \BibitemOpen
  \bibfield  {author} {\bibinfo {author} {\bibfnamefont {E.}~\bibnamefont
  {Tjoa}},\ }\bibfield  {title} {\bibinfo {title} {Fermi two-atom problem:
  Nonperturbative approach via relativistic quantum information and algebraic
  quantum field theory},\ }\href {https://doi.org/10.1103/PhysRevD.106.045012}
  {\bibfield  {journal} {\bibinfo  {journal} {Phys. Rev. D}\ }\textbf {\bibinfo
  {volume} {106}},\ \bibinfo {pages} {045012} (\bibinfo {year}
  {2022})}\BibitemShut {NoStop}%
\bibitem [{\citenamefont {de~Ram\'on}\ \emph {et~al.}(2023)\citenamefont
  {de~Ram\'on}, \citenamefont {Papageorgiou},\ and\ \citenamefont
  {Mart\'{\i}n-Mart\'{\i}nez}}]{Ramon23}%
  \BibitemOpen
  \bibfield  {author} {\bibinfo {author} {\bibfnamefont {J.}~\bibnamefont
  {de~Ram\'on}}, \bibinfo {author} {\bibfnamefont {M.}~\bibnamefont
  {Papageorgiou}},\ and\ \bibinfo {author} {\bibfnamefont {E.}~\bibnamefont
  {Mart\'{\i}n-Mart\'{\i}nez}},\ }\bibfield  {title} {\bibinfo {title}
  {Causality and signalling in noncompact detector-field interactions},\ }\href
  {https://doi.org/10.1103/PhysRevD.108.045015} {\bibfield  {journal} {\bibinfo
   {journal} {Phys. Rev. D}\ }\textbf {\bibinfo {volume} {108}},\ \bibinfo
  {pages} {045015} (\bibinfo {year} {2023})}\BibitemShut {NoStop}%
\bibitem [{\citenamefont {Polo-G\'omez}\ \emph {et~al.}(2022)\citenamefont
  {Polo-G\'omez}, \citenamefont {Garay},\ and\ \citenamefont
  {Mart\'{\i}n-Mart\'{\i}nez}}]{Polo}%
  \BibitemOpen
  \bibfield  {author} {\bibinfo {author} {\bibfnamefont {J.}~\bibnamefont
  {Polo-G\'omez}}, \bibinfo {author} {\bibfnamefont {L.~J.}\ \bibnamefont
  {Garay}},\ and\ \bibinfo {author} {\bibfnamefont {E.}~\bibnamefont
  {Mart\'{\i}n-Mart\'{\i}nez}},\ }\bibfield  {title} {\bibinfo {title} {A
  detector-based measurement theory for quantum field theory},\ }\href
  {https://doi.org/10.1103/PhysRevD.105.065003} {\bibfield  {journal} {\bibinfo
   {journal} {Phys. Rev. D}\ }\textbf {\bibinfo {volume} {105}},\ \bibinfo
  {pages} {065003} (\bibinfo {year} {2022})}\BibitemShut {NoStop}%
\bibitem [{\citenamefont {Perche}\ \emph {et~al.}(2024)\citenamefont {Perche},
  \citenamefont {Polo-G\'omez}, \citenamefont {Torres},\ and\ \citenamefont
  {Mart\'{\i}n-Mart\'{\i}nez}}]{Perche2024}%
  \BibitemOpen
  \bibfield  {author} {\bibinfo {author} {\bibfnamefont {T.~R.}\ \bibnamefont
  {Perche}}, \bibinfo {author} {\bibfnamefont {J.}~\bibnamefont
  {Polo-G\'omez}}, \bibinfo {author} {\bibfnamefont {B.~d. S.~L.}\ \bibnamefont
  {Torres}},\ and\ \bibinfo {author} {\bibfnamefont {E.}~\bibnamefont
  {Mart\'{\i}n-Mart\'{\i}nez}},\ }\bibfield  {title} {\bibinfo {title}
  {Particle detectors from localized quantum field theories},\ }\href
  {https://doi.org/10.1103/PhysRevD.109.045013} {\bibfield  {journal} {\bibinfo
   {journal} {Phys. Rev. D}\ }\textbf {\bibinfo {volume} {109}},\ \bibinfo
  {pages} {045013} (\bibinfo {year} {2024})}\BibitemShut {NoStop}%
\bibitem [{\citenamefont {Haag}(1996)}]{Haag1996}%
  \BibitemOpen
  \bibfield  {author} {\bibinfo {author} {\bibfnamefont {R.}~\bibnamefont
  {Haag}},\ }\href@noop {} {\emph {\bibinfo {title} {Local Quantum Physics:
  Fields, Particles, Algebras}}},\ \bibinfo {edition} {2nd}\ ed.\ (\bibinfo
  {publisher} {Springer},\ \bibinfo {year} {1996})\BibitemShut {NoStop}%
\bibitem [{\citenamefont {Sewell}(1982)}]{Sewell1982}%
  \BibitemOpen
  \bibfield  {author} {\bibinfo {author} {\bibfnamefont {G.~L.}\ \bibnamefont
  {Sewell}},\ }\bibfield  {title} {\bibinfo {title} {Quantum fields on
  manifolds: Pct and gravitationally induced thermal states},\ }\href
  {https://doi.org/10.1016/0003-4916(82)90065-1} {\bibfield  {journal}
  {\bibinfo  {journal} {Annals of Physics}\ }\textbf {\bibinfo {volume}
  {141}},\ \bibinfo {pages} {201} (\bibinfo {year} {1982})}\BibitemShut
  {NoStop}%
\bibitem [{\citenamefont {Kay}\ and\ \citenamefont {Wald}(1991)}]{Kay1991}%
  \BibitemOpen
  \bibfield  {author} {\bibinfo {author} {\bibfnamefont {B.~S.}\ \bibnamefont
  {Kay}}\ and\ \bibinfo {author} {\bibfnamefont {R.~M.}\ \bibnamefont {Wald}},\
  }\bibfield  {title} {\bibinfo {title} {Theorems on the uniqueness and thermal
  properties of stationary, nonsingular, quasifree states on spacetimes with a
  bifurcate killing horizon},\ }\href
  {https://doi.org/https://doi.org/10.1016/0370-1573(91)90015-E} {\bibfield
  {journal} {\bibinfo  {journal} {Phys. Rep.}\ }\textbf {\bibinfo {volume}
  {207}},\ \bibinfo {pages} {49} (\bibinfo {year} {1991})}\BibitemShut
  {NoStop}%
\bibitem [{\citenamefont {Christopher J~Fewster}\ and\ \citenamefont
  {Louko}(2016)}]{Fewster}%
  \BibitemOpen
  \bibfield  {author} {\bibinfo {author} {\bibfnamefont {B.~A. J.-A.}\
  \bibnamefont {Christopher J~Fewster}}\ and\ \bibinfo {author} {\bibfnamefont
  {J.}~\bibnamefont {Louko}},\ }\bibfield  {title} {\bibinfo {title} {Waiting
  for unruh},\ }\href
  {https://iopscience.iop.org/article/10.1088/0264-9381/33/16/165003}
  {\bibfield  {journal} {\bibinfo  {journal} {Class. Quantum Grav.}\ }\textbf
  {\bibinfo {volume} {33}},\ \bibinfo {pages} {165003} (\bibinfo {year}
  {2016})}\BibitemShut {NoStop}%
\bibitem [{\citenamefont {Garay}\ \emph {et~al.}(2016)\citenamefont {Garay},
  \citenamefont {Mart\'{\i}n-Mart\'{\i}nez},\ and\ \citenamefont
  {de~Ram\'on}}]{Garay}%
  \BibitemOpen
  \bibfield  {author} {\bibinfo {author} {\bibfnamefont {L.~J.}\ \bibnamefont
  {Garay}}, \bibinfo {author} {\bibfnamefont {E.}~\bibnamefont
  {Mart\'{\i}n-Mart\'{\i}nez}},\ and\ \bibinfo {author} {\bibfnamefont
  {J.}~\bibnamefont {de~Ram\'on}},\ }\bibfield  {title} {\bibinfo {title}
  {Thermalization of particle detectors: The unruh effect and its reverse},\
  }\href {https://doi.org/10.1103/PhysRevD.94.104048} {\bibfield  {journal}
  {\bibinfo  {journal} {Phys. Rev. D}\ }\textbf {\bibinfo {volume} {94}},\
  \bibinfo {pages} {104048} (\bibinfo {year} {2016})}\BibitemShut {NoStop}%
\bibitem [{\citenamefont {Ju\'arez-Aubry}\ and\ \citenamefont
  {Moustos}(2019)}]{Aubry}%
  \BibitemOpen
  \bibfield  {author} {\bibinfo {author} {\bibfnamefont {B.~A.}\ \bibnamefont
  {Ju\'arez-Aubry}}\ and\ \bibinfo {author} {\bibfnamefont {D.}~\bibnamefont
  {Moustos}},\ }\bibfield  {title} {\bibinfo {title} {Asymptotic states for
  stationary unruh-dewitt detectors},\ }\href
  {https://doi.org/10.1103/PhysRevD.100.025018} {\bibfield  {journal} {\bibinfo
   {journal} {Phys. Rev. D}\ }\textbf {\bibinfo {volume} {100}},\ \bibinfo
  {pages} {025018} (\bibinfo {year} {2019})}\BibitemShut {NoStop}%
\bibitem [{\citenamefont {Perche}(2021)}]{Perche2021}%
  \BibitemOpen
  \bibfield  {author} {\bibinfo {author} {\bibfnamefont {T.~R.}\ \bibnamefont
  {Perche}},\ }\bibfield  {title} {\bibinfo {title} {General features of the
  thermalization of particle detectors and the unruh effect},\ }\href
  {https://doi.org/10.1103/PhysRevD.104.065001} {\bibfield  {journal} {\bibinfo
   {journal} {Phys. Rev. D}\ }\textbf {\bibinfo {volume} {104}},\ \bibinfo
  {pages} {065001} (\bibinfo {year} {2021})}\BibitemShut {NoStop}%
\bibitem [{\citenamefont {Ortega}\ \emph {et~al.}(2019)\citenamefont {Ortega},
  \citenamefont {McKay}, \citenamefont {Alhambra},\ and\ \citenamefont
  {Mart\'{\i}n-Mart\'{\i}nez}}]{Ortega}%
  \BibitemOpen
  \bibfield  {author} {\bibinfo {author} {\bibfnamefont {A.}~\bibnamefont
  {Ortega}}, \bibinfo {author} {\bibfnamefont {E.}~\bibnamefont {McKay}},
  \bibinfo {author} {\bibfnamefont {A.~M.}\ \bibnamefont {Alhambra}},\ and\
  \bibinfo {author} {\bibfnamefont {E.}~\bibnamefont
  {Mart\'{\i}n-Mart\'{\i}nez}},\ }\bibfield  {title} {\bibinfo {title} {Work
  distributions on quantum fields},\ }\href
  {https://doi.org/10.1103/PhysRevLett.122.240604} {\bibfield  {journal}
  {\bibinfo  {journal} {Phys. Rev. Lett.}\ }\textbf {\bibinfo {volume} {122}},\
  \bibinfo {pages} {240604} (\bibinfo {year} {2019})}\BibitemShut {NoStop}%
\bibitem [{\citenamefont {Teixid\'o-Bonfill}\ \emph {et~al.}(2020)\citenamefont
  {Teixid\'o-Bonfill}, \citenamefont {Ortega},\ and\ \citenamefont
  {Mart\'{\i}n-Mart\'{\i}nez}}]{Bonfill}%
  \BibitemOpen
  \bibfield  {author} {\bibinfo {author} {\bibfnamefont {A.}~\bibnamefont
  {Teixid\'o-Bonfill}}, \bibinfo {author} {\bibfnamefont {A.}~\bibnamefont
  {Ortega}},\ and\ \bibinfo {author} {\bibfnamefont {E.}~\bibnamefont
  {Mart\'{\i}n-Mart\'{\i}nez}},\ }\bibfield  {title} {\bibinfo {title} {First
  law of quantum field thermodynamics},\ }\href
  {https://doi.org/10.1103/PhysRevA.102.052219} {\bibfield  {journal} {\bibinfo
   {journal} {Phys. Rev. A}\ }\textbf {\bibinfo {volume} {102}},\ \bibinfo
  {pages} {052219} (\bibinfo {year} {2020})}\BibitemShut {NoStop}%
\bibitem [{\citenamefont {Mottola}(1986)}]{Mottola}%
  \BibitemOpen
  \bibfield  {author} {\bibinfo {author} {\bibfnamefont {E.}~\bibnamefont
  {Mottola}},\ }\bibfield  {title} {\bibinfo {title} {Quantum
  fluctuation-dissipation theorem for general relativity},\ }\href
  {https://doi.org/10.1103/PhysRevD.33.2136} {\bibfield  {journal} {\bibinfo
  {journal} {Phys. Rev. D}\ }\textbf {\bibinfo {volume} {33}},\ \bibinfo
  {pages} {2136} (\bibinfo {year} {1986})}\BibitemShut {NoStop}%
\bibitem [{\citenamefont {Iso}\ \emph {et~al.}(2011)\citenamefont {Iso},
  \citenamefont {Okazawa},\ and\ \citenamefont {Zhang}}]{Iso2011}%
  \BibitemOpen
  \bibfield  {author} {\bibinfo {author} {\bibfnamefont {S.}~\bibnamefont
  {Iso}}, \bibinfo {author} {\bibfnamefont {S.}~\bibnamefont {Okazawa}},\ and\
  \bibinfo {author} {\bibfnamefont {S.}~\bibnamefont {Zhang}},\ }\bibfield
  {title} {\bibinfo {title} {Non-equilibrium fluctuations of black hole
  horizons and the generalized second law},\ }\href
  {https://doi.org/https://doi.org/10.10/16/j.physletb.2011.09.114} {\bibfield
  {journal} {\bibinfo  {journal} {Phys. Lett. B}\ }\textbf {\bibinfo {volume}
  {705}},\ \bibinfo {pages} {152} (\bibinfo {year} {2011})}\BibitemShut
  {NoStop}%
\bibitem [{\citenamefont {Bekenstein}(1974)}]{Bekenstein74}%
  \BibitemOpen
  \bibfield  {author} {\bibinfo {author} {\bibfnamefont {J.~D.}\ \bibnamefont
  {Bekenstein}},\ }\bibfield  {title} {\bibinfo {title} {Generalized second law
  of thermodynamics in black-hole physics},\ }\href
  {https://doi.org/10.1103/PhysRevD.9.3292} {\bibfield  {journal} {\bibinfo
  {journal} {Phys. Rev. D}\ }\textbf {\bibinfo {volume} {9}},\ \bibinfo {pages}
  {3292} (\bibinfo {year} {1974})}\BibitemShut {NoStop}%
\bibitem [{\citenamefont {Wall}(2012)}]{Wall2012}%
  \BibitemOpen
  \bibfield  {author} {\bibinfo {author} {\bibfnamefont {A.~C.}\ \bibnamefont
  {Wall}},\ }\bibfield  {title} {\bibinfo {title} {Proof of the generalized
  second law for rapidly changing fields and arbitrary horizon slices},\ }\href
  {https://doi.org/10.1103/PhysRevD.85.104049} {\bibfield  {journal} {\bibinfo
  {journal} {Phys. Rev. D}\ }\textbf {\bibinfo {volume} {85}},\ \bibinfo
  {pages} {104049} (\bibinfo {year} {2012})}\BibitemShut {NoStop}%
\bibitem [{\citenamefont {Liu}\ \emph {et~al.}(2016)\citenamefont {Liu},
  \citenamefont {Goold}, \citenamefont {Fuentes}, \citenamefont {Vedral},
  \citenamefont {Modi},\ and\ \citenamefont {Bruschi}}]{Liu2016}%
  \BibitemOpen
  \bibfield  {author} {\bibinfo {author} {\bibfnamefont {N.}~\bibnamefont
  {Liu}}, \bibinfo {author} {\bibfnamefont {J.}~\bibnamefont {Goold}}, \bibinfo
  {author} {\bibfnamefont {I.}~\bibnamefont {Fuentes}}, \bibinfo {author}
  {\bibfnamefont {V.}~\bibnamefont {Vedral}}, \bibinfo {author} {\bibfnamefont
  {K.}~\bibnamefont {Modi}},\ and\ \bibinfo {author} {\bibfnamefont {D.~E.}\
  \bibnamefont {Bruschi}},\ }\bibfield  {title} {\bibinfo {title} {Quantum
  thermodynamics for a model of an expanding universe},\ }\href
  {https://doi.org/10.1088/0264-9381/33/3/035003} {\bibfield  {journal}
  {\bibinfo  {journal} {Class. Quantum Grav.}\ }\textbf {\bibinfo {volume}
  {33}},\ \bibinfo {pages} {035003} (\bibinfo {year} {2016})}\BibitemShut
  {NoStop}%
\bibitem [{\citenamefont {Basso}\ \emph {et~al.}(2025)\citenamefont {Basso},
  \citenamefont {Maziero},\ and\ \citenamefont {C\'{e}leri}}]{Basso2025}%
  \BibitemOpen
  \bibfield  {author} {\bibinfo {author} {\bibfnamefont {M.~L.~W.}\
  \bibnamefont {Basso}}, \bibinfo {author} {\bibfnamefont {J.}~\bibnamefont
  {Maziero}},\ and\ \bibinfo {author} {\bibfnamefont {L.~C.}\ \bibnamefont
  {C\'{e}leri}},\ }\bibfield  {title} {\bibinfo {title} {Quantum detailed
  fluctuation theorem in curved spacetimes: The observer dependent nature of
  entropy production},\ }\href {https://doi.org/10.1103/PhysRevLett.134.050406}
  {\bibfield  {journal} {\bibinfo  {journal} {Phys. Rev. Lett.}\ }\textbf
  {\bibinfo {volume} {134}},\ \bibinfo {pages} {050406} (\bibinfo {year}
  {2025})}\BibitemShut {NoStop}%
\bibitem [{\citenamefont {Basso}\ \emph {et~al.}(2023)\citenamefont {Basso},
  \citenamefont {Maziero},\ and\ \citenamefont {C\'{e}leri}}]{Basso2023}%
  \BibitemOpen
  \bibfield  {author} {\bibinfo {author} {\bibfnamefont {M.~L.~W.}\
  \bibnamefont {Basso}}, \bibinfo {author} {\bibfnamefont {J.}~\bibnamefont
  {Maziero}},\ and\ \bibinfo {author} {\bibfnamefont {L.~C.}\ \bibnamefont
  {C\'{e}leri}},\ }\bibfield  {title} {\bibinfo {title} {The irreversibility of
  relativistic time-dilation},\ }\href
  {https://iopscience.iop.org/article/10.1088/1361-6382/acf089} {\bibfield
  {journal} {\bibinfo  {journal} {Class. Quantum Grav.}\ }\textbf {\bibinfo
  {volume} {40}},\ \bibinfo {pages} {195001} (\bibinfo {year}
  {2023})}\BibitemShut {NoStop}%
\bibitem [{\citenamefont {Cai}\ \emph {et~al.}(2025)\citenamefont {Cai},
  \citenamefont {Wang},\ and\ \citenamefont {Zhao}}]{Cai25}%
  \BibitemOpen
  \bibfield  {author} {\bibinfo {author} {\bibfnamefont {Y.}~\bibnamefont
  {Cai}}, \bibinfo {author} {\bibfnamefont {T.}~\bibnamefont {Wang}},\ and\
  \bibinfo {author} {\bibfnamefont {L.}~\bibnamefont {Zhao}},\ }\bibfield
  {title} {\bibinfo {title} {General relativistic fluctuation theorems},\
  }\href {https://doi.org/https://doi.org/10.1016/j.physletb.2024.139220}
  {\bibfield  {journal} {\bibinfo  {journal} {Phys. Lett. B}\ }\textbf
  {\bibinfo {volume} {860}},\ \bibinfo {pages} {139220} (\bibinfo {year}
  {2025})}\BibitemShut {NoStop}%
\bibitem [{\citenamefont {Kubo}(1957)}]{Kubo1957}%
  \BibitemOpen
  \bibfield  {author} {\bibinfo {author} {\bibfnamefont {R.}~\bibnamefont
  {Kubo}},\ }\bibfield  {title} {\bibinfo {title} {Statistical-mechanical
  theory of irreversible processes. i. general theory and simple applications
  to magnetic and conduction problems},\ }\href
  {https://doi.org/10.1143/JPSJ.12.570} {\bibfield  {journal} {\bibinfo
  {journal} {J. Phys. Soc. Jpn.}\ }\textbf {\bibinfo {volume} {12}},\ \bibinfo
  {pages} {570} (\bibinfo {year} {1957})}\BibitemShut {NoStop}%
\bibitem [{\citenamefont {Martin}\ and\ \citenamefont
  {Schwinger}(1959)}]{Martin1959}%
  \BibitemOpen
  \bibfield  {author} {\bibinfo {author} {\bibfnamefont {P.~C.}\ \bibnamefont
  {Martin}}\ and\ \bibinfo {author} {\bibfnamefont {J.}~\bibnamefont
  {Schwinger}},\ }\bibfield  {title} {\bibinfo {title} {Theory of many-particle
  systems. i},\ }\href {https://doi.org/10.1103/PhysRev.115.1342} {\bibfield
  {journal} {\bibinfo  {journal} {Phys. Rev.}\ }\textbf {\bibinfo {volume}
  {115}},\ \bibinfo {pages} {1342} (\bibinfo {year} {1959})}\BibitemShut
  {NoStop}%
\bibitem [{\citenamefont {Sanders}(2013)}]{Sanders10}%
  \BibitemOpen
  \bibfield  {author} {\bibinfo {author} {\bibfnamefont {K.}~\bibnamefont
  {Sanders}},\ }\bibfield  {title} {\bibinfo {title} {Thermal equilibrium
  states of a linear scalar quantum field in stationary space–times},\ }\href
  {https://doi.org/10.1142/S0217751X1330010X} {\bibfield  {journal} {\bibinfo
  {journal} {Int. J. Mod. Phys. A}\ }\textbf {\bibinfo {volume} {28}},\
  \bibinfo {pages} {1330010} (\bibinfo {year} {2013})}\BibitemShut {NoStop}%
\bibitem [{\citenamefont {Perche}(2022)}]{perche2022localized}%
  \BibitemOpen
  \bibfield  {author} {\bibinfo {author} {\bibfnamefont {T.~R.}\ \bibnamefont
  {Perche}},\ }\bibfield  {title} {\bibinfo {title} {Localized non-relativistic
  quantum systems in curved spacetimes: a general characterization of particle
  detector models},\ }\href {https://doi.org/10.1103/PhysRevD.106.025018}
  {\bibfield  {journal} {\bibinfo  {journal} {Phys. Rev. D}\ }\textbf {\bibinfo
  {volume} {106}},\ \bibinfo {pages} {025018} (\bibinfo {year}
  {2022})}\BibitemShut {NoStop}%
\bibitem [{\citenamefont {Manasse}\ and\ \citenamefont
  {Misner}(1963)}]{manasse1963fermi}%
  \BibitemOpen
  \bibfield  {author} {\bibinfo {author} {\bibfnamefont {F.~K.}\ \bibnamefont
  {Manasse}}\ and\ \bibinfo {author} {\bibfnamefont {C.~W.}\ \bibnamefont
  {Misner}},\ }\bibfield  {title} {\bibinfo {title} {Fermi normal coordinates
  and some basic concepts in differential geometry},\ }\href
  {https://doi.org/10.1063/1.1724316} {\bibfield  {journal} {\bibinfo
  {journal} {J. Math. Phys.}\ }\textbf {\bibinfo {volume} {4}},\ \bibinfo
  {pages} {735} (\bibinfo {year} {1963})}\BibitemShut {NoStop}%
\bibitem [{\citenamefont {Poisson}(2004)}]{poisson2004relativist}%
  \BibitemOpen
  \bibfield  {author} {\bibinfo {author} {\bibfnamefont {E.}~\bibnamefont
  {Poisson}},\ }\href@noop {} {\emph {\bibinfo {title} {A relativist's toolkit:
  The mathematics of black-hole mechanics}}}\ (\bibinfo  {publisher} {Cambridge
  University Press},\ \bibinfo {year} {2004})\BibitemShut {NoStop}%
\bibitem [{\citenamefont {Medved}\ \emph {et~al.}(2004)\citenamefont {Medved},
  \citenamefont {Martin},\ and\ \citenamefont {Visser}}]{Medved04}%
  \BibitemOpen
  \bibfield  {author} {\bibinfo {author} {\bibfnamefont {A.~J.~M.}\
  \bibnamefont {Medved}}, \bibinfo {author} {\bibfnamefont {D.}~\bibnamefont
  {Martin}},\ and\ \bibinfo {author} {\bibfnamefont {M.}~\bibnamefont
  {Visser}},\ }\bibfield  {title} {\bibinfo {title} {Dirty black holes:
  spacetime geometry and near-horizon symmetries},\ }\href
  {https://doi.org/10.1088/0264-9381/21/13/003} {\bibfield  {journal} {\bibinfo
   {journal} {Class. and Quantum Grav.}\ }\textbf {\bibinfo {volume} {21}},\
  \bibinfo {pages} {3111} (\bibinfo {year} {2004})}\BibitemShut {NoStop}%
\bibitem [{\citenamefont {Mart\'{\i}n-Mart\'{\i}nez}\ \emph
  {et~al.}(2020)\citenamefont {Mart\'{\i}n-Mart\'{\i}nez}, \citenamefont
  {Perche},\ and\ \citenamefont {de~S.~L.~Torres}}]{Martinez20}%
  \BibitemOpen
  \bibfield  {author} {\bibinfo {author} {\bibfnamefont {E.}~\bibnamefont
  {Mart\'{\i}n-Mart\'{\i}nez}}, \bibinfo {author} {\bibfnamefont {T.~R.}\
  \bibnamefont {Perche}},\ and\ \bibinfo {author} {\bibfnamefont
  {B.}~\bibnamefont {de~S.~L.~Torres}},\ }\bibfield  {title} {\bibinfo {title}
  {General relativistic quantum optics: Finite-size particle detector models in
  curved spacetimes},\ }\href {https://doi.org/10.1103/PhysRevD.101.045017}
  {\bibfield  {journal} {\bibinfo  {journal} {Phys. Rev. D}\ }\textbf {\bibinfo
  {volume} {101}},\ \bibinfo {pages} {045017} (\bibinfo {year}
  {2020})}\BibitemShut {NoStop}%
\bibitem [{\citenamefont {Mart\'{\i}n-Mart\'{\i}nez}\ \emph
  {et~al.}(2021)\citenamefont {Mart\'{\i}n-Mart\'{\i}nez}, \citenamefont
  {Perche},\ and\ \citenamefont {Torres}}]{Martinez21}%
  \BibitemOpen
  \bibfield  {author} {\bibinfo {author} {\bibfnamefont {E.}~\bibnamefont
  {Mart\'{\i}n-Mart\'{\i}nez}}, \bibinfo {author} {\bibfnamefont {T.~R.}\
  \bibnamefont {Perche}},\ and\ \bibinfo {author} {\bibfnamefont {B.~d. S.~L.}\
  \bibnamefont {Torres}},\ }\bibfield  {title} {\bibinfo {title} {Broken
  covariance of particle detector models in relativistic quantum information},\
  }\href {https://doi.org/10.1103/PhysRevD.103.025007} {\bibfield  {journal}
  {\bibinfo  {journal} {Phys. Rev. D}\ }\textbf {\bibinfo {volume} {103}},\
  \bibinfo {pages} {025007} (\bibinfo {year} {2021})}\BibitemShut {NoStop}%
\bibitem [{\citenamefont {Blanes}\ \emph {et~al.}(2009)\citenamefont {Blanes},
  \citenamefont {Casas}, \citenamefont {Oteo},\ and\ \citenamefont
  {Ros}}]{Blanes09}%
  \BibitemOpen
  \bibfield  {author} {\bibinfo {author} {\bibfnamefont {S.}~\bibnamefont
  {Blanes}}, \bibinfo {author} {\bibfnamefont {F.}~\bibnamefont {Casas}},
  \bibinfo {author} {\bibfnamefont {J.}~\bibnamefont {Oteo}},\ and\ \bibinfo
  {author} {\bibfnamefont {J.}~\bibnamefont {Ros}},\ }\bibfield  {title}
  {\bibinfo {title} {The magnus expansion and some of its applications},\
  }\href {https://doi.org/https://doi.org/10.1016/j.physrep.2008.11.001}
  {\bibfield  {journal} {\bibinfo  {journal} {Phys. Rep.}\ }\textbf {\bibinfo
  {volume} {470}},\ \bibinfo {pages} {151} (\bibinfo {year}
  {2009})}\BibitemShut {NoStop}%
\bibitem [{\citenamefont {Landulfo}(2016)}]{Landulfo}%
  \BibitemOpen
  \bibfield  {author} {\bibinfo {author} {\bibfnamefont {A.~G.~S.}\
  \bibnamefont {Landulfo}},\ }\bibfield  {title} {\bibinfo {title}
  {Nonperturbative approach to relativistic quantum communication channels},\
  }\href {https://doi.org/10.1103/PhysRevD.93.104019} {\bibfield  {journal}
  {\bibinfo  {journal} {Phys. Rev. D}\ }\textbf {\bibinfo {volume} {93}},\
  \bibinfo {pages} {104019} (\bibinfo {year} {2016})}\BibitemShut {NoStop}%
\bibitem [{\citenamefont {Jarzynski}(2011)}]{Jarzynski11}%
  \BibitemOpen
  \bibfield  {author} {\bibinfo {author} {\bibfnamefont {C.}~\bibnamefont
  {Jarzynski}},\ }\bibfield  {title} {\bibinfo {title} {Equalities and
  inequalities: Irreversibility and the second law of thermodynamics at the
  nanoscale},\ }\href
  {https://doi.org/https://doi.org/10.1146/annurev-conmatphys-062910-140506}
  {\bibfield  {journal} {\bibinfo  {journal} {Annual Review of Condensed Matter
  Physics}\ }\textbf {\bibinfo {volume} {2}},\ \bibinfo {pages} {329} (\bibinfo
  {year} {2011})}\BibitemShut {NoStop}%
\bibitem [{\citenamefont {Poisson}\ \emph {et~al.}(2011)\citenamefont
  {Poisson}, \citenamefont {Pound},\ and\ \citenamefont {Vega}}]{Poisson11}%
  \BibitemOpen
  \bibfield  {author} {\bibinfo {author} {\bibfnamefont {E.}~\bibnamefont
  {Poisson}}, \bibinfo {author} {\bibfnamefont {A.}~\bibnamefont {Pound}},\
  and\ \bibinfo {author} {\bibfnamefont {I.}~\bibnamefont {Vega}},\ }\bibfield
  {title} {\bibinfo {title} {The motion of point particles in curved
  spacetime},\ }\href {https://doi.org/10.12942/lrr-2011-7} {\bibfield
  {journal} {\bibinfo  {journal} {Living Rev. Relativ.}\ }\textbf {\bibinfo
  {volume} {14}},\ \bibinfo {pages} {7} (\bibinfo {year} {2011})}\BibitemShut
  {NoStop}%
\bibitem [{\citenamefont {Solveen}(2012)}]{Solveen12}%
  \BibitemOpen
  \bibfield  {author} {\bibinfo {author} {\bibfnamefont {C.}~\bibnamefont
  {Solveen}},\ }\bibfield  {title} {\bibinfo {title} {Local thermal equilibrium
  and kms states in curved spacetime},\ }\href
  {https://doi.org/10.1088/0264-9381/29/24/245015} {\bibfield  {journal}
  {\bibinfo  {journal} {Class. Quantum Grav.}\ }\textbf {\bibinfo {volume}
  {29}},\ \bibinfo {pages} {245015} (\bibinfo {year} {2012})}\BibitemShut
  {NoStop}%
\bibitem [{\citenamefont {Gransee}\ \emph {et~al.}(2017)\citenamefont
  {Gransee}, \citenamefont {Pinamonti},\ and\ \citenamefont
  {Verch}}]{Gransee17}%
  \BibitemOpen
  \bibfield  {author} {\bibinfo {author} {\bibfnamefont {M.}~\bibnamefont
  {Gransee}}, \bibinfo {author} {\bibfnamefont {N.}~\bibnamefont {Pinamonti}},\
  and\ \bibinfo {author} {\bibfnamefont {R.}~\bibnamefont {Verch}},\ }\bibfield
   {title} {\bibinfo {title} {Kms-like properties of local equilibrium states
  in quantum field theory},\ }\href
  {https://doi.org/https://doi.org/10.1016/j.geomphys.2017.02.014} {\bibfield
  {journal} {\bibinfo  {journal} {J. Geom. Phys.}\ }\textbf {\bibinfo {volume}
  {117}},\ \bibinfo {pages} {15} (\bibinfo {year} {2017})}\BibitemShut
  {NoStop}%
\bibitem [{\citenamefont {Wald}(2001)}]{Wald01}%
  \BibitemOpen
  \bibfield  {author} {\bibinfo {author} {\bibfnamefont {R.}~\bibnamefont
  {Wald}},\ }\bibfield  {title} {\bibinfo {title} {The thermodynamics of black
  holes},\ }\href {https://doi.org/10.12942/lrr-2001-6} {\bibfield  {journal}
  {\bibinfo  {journal} {Living Rev. Relativ.}\ }\textbf {\bibinfo {volume}
  {4}},\ \bibinfo {pages} {6} (\bibinfo {year} {2001})}\BibitemShut {NoStop}%
\bibitem [{\citenamefont {Marolf}\ \emph {et~al.}(2004)\citenamefont {Marolf},
  \citenamefont {Minic},\ and\ \citenamefont {Ross}}]{Marolf04}%
  \BibitemOpen
  \bibfield  {author} {\bibinfo {author} {\bibfnamefont {D.}~\bibnamefont
  {Marolf}}, \bibinfo {author} {\bibfnamefont {D.}~\bibnamefont {Minic}},\ and\
  \bibinfo {author} {\bibfnamefont {S.~F.}\ \bibnamefont {Ross}},\ }\bibfield
  {title} {\bibinfo {title} {Notes on spacetime thermodynamics and the observer
  dependence of entropy},\ }\href {https://doi.org/10.1103/PhysRevD.69.064006}
  {\bibfield  {journal} {\bibinfo  {journal} {Phys. Rev. D}\ }\textbf {\bibinfo
  {volume} {69}},\ \bibinfo {pages} {064006} (\bibinfo {year}
  {2004})}\BibitemShut {NoStop}%
\bibitem [{\citenamefont {Oliveira}\ and\ \citenamefont
  {Céleri}(2024)}]{Oliveira2024}%
  \BibitemOpen
  \bibfield  {author} {\bibinfo {author} {\bibfnamefont {G.}~\bibnamefont
  {Oliveira}}\ and\ \bibinfo {author} {\bibfnamefont {L.~C.}\ \bibnamefont
  {Céleri}},\ }\bibfield  {title} {\bibinfo {title} {Thermodynamic entropy
  production in the dynamical casimir effect},\ }\href
  {https://doi.org/10.1103/PhysRevA.109.012807} {\bibfield  {journal} {\bibinfo
   {journal} {Phys. Rev. A}\ }\textbf {\bibinfo {volume} {109}},\ \bibinfo
  {pages} {012807} (\bibinfo {year} {2024})}\BibitemShut {NoStop}%
\bibitem [{\citenamefont {Moreira}\ and\ \citenamefont
  {Céleri}(2025)}]{Moreira2025}%
  \BibitemOpen
  \bibfield  {author} {\bibinfo {author} {\bibfnamefont {T.~H.}\ \bibnamefont
  {Moreira}}\ and\ \bibinfo {author} {\bibfnamefont {L.~C.}\ \bibnamefont
  {Céleri}},\ }\bibfield  {title} {\bibinfo {title} {Entropy production due to
  spacetime fluctuations},\ }\href
  {https://iopscience.iop.org/article/10.1088/1361-6382/ada083} {\bibfield
  {journal} {\bibinfo  {journal} {Class. Quantum Grav.}\ }\textbf {\bibinfo
  {volume} {42}},\ \bibinfo {pages} {025022} (\bibinfo {year}
  {2025})}\BibitemShut {NoStop}%
\end{thebibliography}
\end{document}